\documentclass{aa}
\usepackage{hyperref}
\usepackage[varg]{txfonts}
\usepackage[utf8]{inputenc}
\usepackage{amsmath}
\usepackage{subfig}
\usepackage{blindtext}
\usepackage{makecell} 
\usepackage[normalem]{ulem}
\usepackage{color}
\usepackage{multirow}
\definecolor{darkgreen}{rgb}{0.0,0.5,0.0}
\definecolor{darkred}{rgb}{0.75,0.,0.2}
\definecolor{magenta}{rgb}{0.8,0,0.8}
\definecolor{purple}{rgb}{0.5,0,0.5}
\definecolor{gray}{rgb}{0.5,0.6,0.7}
\definecolor{orange}{rgb}{1.0,0.3,0.0}
\definecolor{Black}{rgb}{0,0,0}
\newcommand\redsout{\bgroup\markoverwith{\textcolor{orange}{\rule[0.5ex]{2pt}{0.8pt}}}\ULon}

\hypersetup{
  colorlinks = true,
  urlcolor   = blue,
  linkcolor  = blue,
  citecolor  = blue
}
\usepackage{amssymb}
\usepackage{natbib}
\usepackage{journals}
\bibpunct{(}{)}{;}{a}{}{,} 
\setcitestyle{aysep={},yysep={;}}

\newcommand{\Compact}{Compact}
\newcommand{\Compacts}{Compacts}
\newcommand{\CompactMB}{Compact$_\mathrm{MB}$}
\newcommand{\CompactsMB}{Compacts$_\mathrm{MB}$}
\newcommand{\CompactMBs}{Compacts$_\mathrm{MB}$}
\newcommand{\CompactSB}{Compact$_\mathrm{SB}$}
\newcommand{\CompactsSB}{Compacts$_\mathrm{SB}$}
\newcommand{\CompactSBs}{Compacts$_\mathrm{SB}$}
\newcommand{\Normal}{Normal}
\newcommand{\Normals}{Normals}

\newcommand{\msun}{\,\mathrm{M}_\odot}
\newcommand{\lmsun}{\log(M_\star/{\rm M}_\odot)}
\usepackage{upgreek}
\renewcommand{\pi}{\uppi}
\begin{document}
    
        \title{What drives the corpulence of galaxies? I. The formation of central compact dwarf galaxies in TNG50}
 \titlerunning{Formation of compact central dwarf galaxies}
\author{Abhner P. de Almeida
                 \inst{1,2}\fnmsep\thanks{\email{abhner.almeida@usp.br}}
                 \and
                 Gary A. Mamon\inst{1}
        \and
                 Avishai Dekel \inst{3}
                 \and
        Gastão B. Lima Neto\inst{2}
                 }

   \institute{Institut d'Astrophysique de Paris (UMR 7095: CNRS \& Sorbonne Universit\'e), 98 bis Bd Arago, F-75014 Paris, France 
                 \and
                 Instituto de Astronomia, Geof\'isica e Ci\^encias Atmosf\'ericas (Universidade de S\~ao Paulo), R. do Mat\~ao, 1226, S\~ao Paulo - SP, 05508-090, Brasil
    \and 
    Racah Institute of Physics, The Hebrew University, Jerusalem 91904, Israel
                 }

   \date{Received Month, day, year; accepted Month day, year}

 
  \abstract{
      Nearby dwarf galaxies display a variety of effective radii (sizes)  at a given stellar mass, suggesting different evolution scenarios according to their final `stellar' size. The  TNG hydrodynamical  simulations present a  bimodality in the $z$=0 size--mass relation (SMRz0) of dwarf galaxies,  at $r_{1/2,\star}\sim$450 pc.   
         Using the TNG50 simulation, we explored the evolution of the most massive progenitors of dwarf galaxies ($z$=0 $\lmsun$ between 8.4 and 9.2) that end up as central galaxies of their groups. We split these dwarfs into three classes of the SMRz0: ``\Normals\ '' from the central spine of the main branch, and ``\Compacts\ '' from the secondary branch as well as the lower envelope of the main branch.
    Both classes of \Compacts\ see their stellar sizes decrease from $z$$\sim$1 onwards in contrast to \Normals, while  the sizes of the gas and dark matter (DM) components continue to increase (as for \Normals). 
  A detailed analysis reveals that \Compacts\ live in poorer environments, and thus suffer fewer major mergers from $z=0.8$ onwards, which otherwise would pump angular momentum into the gas, allowing strong gas inflows, producing inner star formation, and thus leading to the buildup of a stellar core. \Compacts\ are predicted to be rounder and to have bluer cores. Compact dwarfs of similar sizes are observed in the GAMA survey, but the bimodality in size is less evident and the most compact dwarfs tend to be passive rather than star forming, as in TNG50. Our conclusions should therefore be confirmed with future cosmological hydrodynamical simulations.}
      

%

\maketitle

\section{Introduction}

Dwarf galaxies (hereafter dwarfs) constitute an important laboratory for studying the influence of different physical mechanisms on the formation and evolution of galaxies. They mainly acquire their stellar mass from star formation driven by gas accretion. Indeed, the other path for star formation, namely  starbursts from gas-rich galaxy mergers, should not be important because low-mass galaxies rarely acquire the bulk of their stellar mass from mergers \citep{Guo&White08,Cattaneo+11,Bernardi+11}.

Dwarfs have difficulty in maintaining the supply of gas from accretion, because of several mechanisms, both internal and external. Internal processes include energetic feedback from supernovae \citep{Dekel1986ApJ} and from the central supermassive black hole \citep{Silk1998}, if present. External processes include a variety of environmental effects, such as collisional tides \citep{Richstone1976ApJ}; tides from  clusters \citep{Merritt1983ApJ} and groups \citep{Mamon87}; ram pressure \citep{Gunn1972ApJ};  harassment from numerous flybys \citep{Moore1996Natur} and the effects of a nearby active galactic nucleus \citep{Dashyan+19}. These external processes mainly  affect galaxies that end up as satellites. Interactions between galaxies can lead to different results depending on how they occur and the galaxies involved \cite[e.g.,][]{Barnes1991ApJ, Barnes1992ARAA}.

One fundamental issue is the size--mass relation (SMR) of galaxies, that is the relation between their {effective} (projected half-light) radius and their stellar mass. Analyzing galaxies from the Main Galaxy Sample of the Sloan Digital Sky Survey (SDSS), \cite{Shen2003MNRAS} found that the SMR of galaxies is not universal: it is steeper for  early-type galaxies than for late-type galaxies, leading to smaller sizes for early-type galaxies once one extrapolates to the low-mass end.

Low-mass ($M_\star$  between $10^7$ and $10^9\,\msun$) galaxies are bimodal in size, with a split at a few hundred parsecs (e.g., \citealt{Misgeld2011MNRAS}), with a diffuse class (dwarf ellipticals and dwarf spheroidals) on one hand and a compact class (compact ellipticals (cEs), ultracompact dwarfs (UCDs), and globular clusters) on the other. The compact class systems have sizes of less than 10 (respectively, 100) times the median sizes of galaxies with $\lmsun=8$ (respectively, 7).\footnote{All our logarithms are in base 10.}

The strong bimodality in the sizes of dwarf galaxies indicate that the scenarios of formation and evolution of compact dwarfs are different from those of high-mass galaxies. For example, the observation of compact dwarf galaxies (both UCDs and cEs) in galaxy clusters or near massive galaxies suggests formation scenarios associated with tidal stripping of larger galaxies, groups, or clusters \cite[e.g.,][]{ Chilingarian&Mamon08, Brodie2011AJ, Wang2023Nature}.

But tidal stripping does not affect central galaxies. Several other physical mechanisms have been suggested to form compact dwarfs. For example, UCDs could be extremely massive globular clusters (\citealp*{Mieske2002AA}; \citealp*{Mieske2012AA}) or the result of mergers of young globular clusters formed in merging galaxies \citep{Kroupa1998MNRAS, Mahani2021MNRAS}. Blue compact dwarfs (BCDs) may be the result of mergers, specifically dwarf--dwarf mergers that can trigger a central starburst \citep[e.g.,][]{Bekki2008MNRAS, Watts2016MNRAS}.

High-mass galaxies can also be compact: \cite{vanDokkum+08} discovered compact $10^{10} \msun$ galaxies at $z=2$, with effective radii of as low as 500~pc, corresponding to  spherical half-mass radii of roughly 750 pc.
Dekel and collaborators (\citealp*{Dekel+09}; \citealp{Dekel2014MNRAS,Zolotov2015MNRAS})  argued that these high-redshift, relatively massive compact galaxies have a three-phase cycle: they form from violent disk instabilities in normal-size  galaxies, causing {compaction} of the gas, which becomes so dense that it creates a compact starburst ({blue nugget}) and later passively evolves into a {red nugget}.  For massive compact galaxies observed in the local Universe (probed by the Mapping Nearby Galaxies at Apache Point Observatory (MaNGA)), \cite{Schnorr2021MNRAS}  suggested an analogous scenario, occurring between $z=2$ and 0.4, albeit less extreme.   In a  followup by the same team on the Illustris TNG100 cosmological hydrodynamical simulation, \cite{Lohmann+23} recently  pointed out that compact galaxies  accrete gas with less angular momentum than that accreted by other galaxies, allowing compact galaxies  to increase their mass without significantly growing in size. In a similar analysis of galaxies with resolved spectroscopy (Sydney-AAO Multi-object Integral field spectrograph, SAMI) and with the Illustris TNG50 simulation, \cite{Deeley+23} found that one-third of compact massive dwarfs were stripped by a massive host galaxy.  These authors also noted that the remaining isolated ones preferentially formed stars in their inner regions, although no physical mechanism differentiating these galaxies from normal ones was given.

The SMRs of galaxies in hydrodynamical simulations are in good agreement with observations for galaxies with $M_{\rm stars} > 10^9 \msun$ \citep{Furlong2017MNRAS, Genel2018MNRAS}.  A striking feature of the Illustris simulations is the existence of a second, small-size branch in the $z$=0 SMR, whose slope is, surprisingly, negative (Fig.~10 of \citealt{Haslbauer+19} for Illustris, upper-left panel of Fig.~4 of \citealt{Genel2018MNRAS} for TNG100, and lower-right panel of Fig.~4 and Fig.~8 of \citealt{Pillepich2018MNRAS} for TNG50).  A closer look using the {\sf Plot Group/Halo    Catalogs} facility on the TNG web site\footnote{https://www.tng-project.org/data/groupcat/} reveals that this negative-slope small-size mode extends to stellar masses of $10^{9.5}$ to $10^{10}\,{\rm M}_{\odot}$ for TNG50, TNG100, and TNG300. This secondary negative-slope branch of the SMR has not yet been explained.

In the present article, we present a study on  the physical mechanisms producing ``low-corpulence'' dwarf galaxies that end up as the central galaxy of their host.  As the SMR extends to different stellar-mass ranges, we use corpulence to refer to the stellar half-mass radius given a specific stellar mass.  We considered both the lower-corpulence galaxies in the main branch of the SMR and the lower-corpulence (secondary) branch of the  SMR. To this end, we used  the well-resolved TNG50-1 simulation, by following the evolution of the main progenitors of $z$=0  galaxies. In two forthcoming studies (de Almeida et al., in prep.), we will use our tools to study the physical mechanisms driving {low-corpulence dwarfs that end up as satellites} as well as those processes leading to {high-corpulence} (highly diffuse) dwarfs.

In Sect.~\ref{sec:sim&sample}, we present the TNG50 simulation as well as  our selection of compact galaxies and a control sample of normal galaxies in the same range of stellar mass. In Sect.~\ref{sec:evol}, we present an analysis of the median evolution of the main progenitors of present-day compact and normal galaxies. In Sect.~\ref{sec:details}, we explore the respective roles of the physical   mechanisms driving size evolution. We discuss our results in Sect.~\ref{Discussion} and summarize our findings in Sect.~\ref{Conclusion}.

\section{Simulation and sample}
\label{sec:sim&sample}

\subsection{IllustrisTNG}
\label{sec:TNG}

We studied the evolution of dwarf compact galaxies using the IllustrisTNG (hereafter, TNG) suite of cosmological magneto-hydrodynamical simulations \citep{Springel2018MNRAS, Pillepich2018bMNRAS, Marinacci2018MNRAS, Naiman2018MNRAS, Nelson2018MNRAS}. These simulations (as in the previous  Illustris simulations) were run with the {\sc arepo} magneto-hydrodynamics moving mesh code \citep{Springel10}. The TNG simulations were run with initial conditions drawn from the cosmological parameters values from \cite{Planck2016AA}: a flat $\Lambda$CDM cosmology with Hubble constant $H_0 = 67.74\, \mathrm{km \,s^{-1} \, Mpc^{-1}}$, matter density $\Omega_{\rm m} = 0.3089$, baryon density $\Omega_{\rm b} = 0.0486$, power spectrum normalization $\sigma_8 = 0.8159$, and primordial spectral index $n_s = 0.9667$.
The TNG suite  has a variety of box sizes and resolutions, and some were run without baryons. For the simulations with baryons, TNG implements prescriptions for star formation, stellar feedback, metal enrichment, and black hole (BH) physics (seeding and feedback). All simulations follow the evolution of a comoving box of the Universe (with periodic boundary conditions) from $z = 127$ to $z = 0$, with data saved at 100 different snapshots, typically spaced by 150 Myr in time.

The TNG database provides tables of groups, subhalos, particles (gas,   dark matter (DM), stars, black holes, as well as tracer particles not considered  here), subhalo merger trees, and 33 (at the time of the writing of this manuscript) additional tables provided by the users. The groups are identified  using the friends-of-friends  algorithm \citep{Davis1985ApJ} on the particles, while the subhalos are the structures within groups identified using the {\sc subfind} algorithm \citep{Springel2001MNRAS}. The subhalos are mostly galaxies and the TNG database provides numerous astrophysical attributes for each one. One can have groups with a single subhalo, representing an isolated galaxy. We hereafter use the term {galaxies} to refer to subhalos and often use {hosts} to refer to groups.

In TNG, a subhalo has a bad flag if it forms  within one virial radius of a group, and with less than 80\% of its mass in DM.  This flag allows users to discard \textup{H\,\textsc{ii}} regions incorrectly extracted by {\sc subfind}. 

\subsection{The TNG50 simulation}
\label{sec:TNG50}

We adopt the TNG50-1 (hereafter, TNG50) simulation  \citep{Nelson2019MNRAS, Pillepich2019MNRAS}, as it is the best resolved and most suited to studying  the evolution of   dwarf galaxies. The DM and gas resolutions of the  simulation are $m_\mathrm{DM} = 4.5 \times 10^5 \msun$ and $m_\mathrm{gas} = 8.5 \times 10^4 \msun$, respectively, in a volume of $51.7^3\, \mathrm{Mpc^3}$. The size resolution is described by  softening lengths of the collisionless components (DM and stars) whose physical values increase with cosmic time at 576 comoving pc until $z=1$, and are then fixed at the $z=1$ physical size of 288 pc \citep{Pillepich2019MNRAS}. The gas softening
length is 74 comoving pc (the minimum gas cell size is 8 pc). Thus, galactic
disks are reasonably well resolved in TNG50.
The TNG50 simulations have the same subgrid physics as the TNG100 and TNG300 simulations, except for a slightly different criterion for the star formation time, which is designed to avoid instantaneous star formation in dense regions \citep{Nelson2019MNRAS}.

\subsection{Sample selection}
\label{sec:sample}

\begin{figure}[ht]
  \centering
  \includegraphics[width=\hsize]{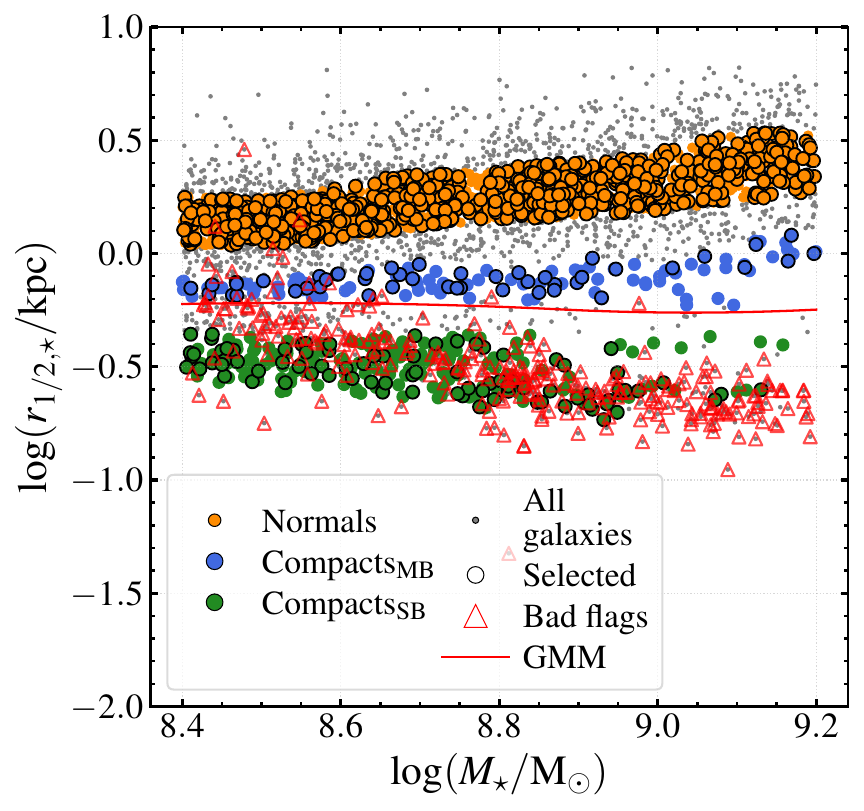}
  \caption{
Present-day stellar half-mass radius vs. stellar mass (within 2 stellar half-mass radii), highlighting our  adopted samples of \Compacts\ and \Normals{} (both centrals and satellites).   The {orange}, {blue,} and {green circles}  are the different samples, respectively: \Normals, \CompactsMB\ (Main branch) and \CompactsSB\ (Secondary branch), while the {gray dots} are all the subhalos in the stellar mass range. The triangles are bad-flag galaxies. The {red solid line} is the separation made with a Gaussian mixing model (GMM). }
   \label{fig:sizevsmass_classes}
   \end{figure}
   
Figure \ref{fig:sizevsmass_classes} displays the present-day size--mass relation of galaxies in our chosen range of stellar masses.  The bimodality of sizes is clear: there are only few small gray points at  $\log(r_{1/2,\star}/{\rm kpc})\approx -0.25$ (i.e., 560 pc); see the red curve found with a   Gaussian mixing model (we hereafter use $\log(r_{1/2}/{\rm kpc})$ for the stellar half-mass radius).\footnote{We used  {\tt sklearn.mixtureGaussianMixture}, considering both good- and bad-flag subhalos.} We denote the large- and small-size branches  ``{Main}'' and ``{Secondary}'', respectively.  The Secondary branch has a negative slope, as \cite{Haslbauer+19} and \cite{Genel2018MNRAS} previously found in Illustris and  TNG100, respectively. As the Secondary branch of the SMR is polluted by suspicious subhalos (one-third with bad flags, and mostly low-DM galaxies), we considered two samples of ``\Compact\ '' galaxies: the lower envelope of the Main branch of the SMR, and the Secondary branch, both restricted to good-flag galaxies. 

We selected all good-flag subhalos with $\log ( M_{\star} / \msun ) > 8.4$ in order to have a sufficient number of particles to resolve the galaxy and be able to efficiently separate the Main and Secondary branches. We also restrict the subhalos to $\log ( M_{\star} / \msun )< 9.2$ in order to limit our selection to  low-mass galaxies and avoid the poorly populated end of the Secondary branch.\footnote{These stellar masses correspond to those enclosed in a sphere of twice the stellar half-mass radius in order to mimic the fact that observers have trouble measuring the outer luminosities (hence stellar masses) of galaxies.} Finally, we only consider subhalos that have an identified progenitor.

We then defined a first set of \Compact\ galaxies taken from a  conservative subset of the Secondary branch: ``\CompactsSB{}'' are the galaxies whose (3D) stellar half-mass radius satisfies $\log(r_{1/2}/{\rm kpc}) < -0.35$ (i.e. $r_{1/2}  < 447\,\rm pc$). We also defined a second set of \Compact\ galaxies, hereafter ``\CompactsMB'', taken from the lower 5th  percentile of the Main branch (taking the galaxy sizes for different stellar mass bins and computing the median and the  5th percentile of the residuals). The minimum and maximum sizes of \CompactsMB{} as a function of stellar mass within $2\,r_{1/2}$ can be approximated as
\begin{equation}
\label{eq:01}
\begin{split}
& \log\left({r_{1/2, \mathrm{\, max}} \over \mathrm{kpc}}\right) = 0.29 \,\log\left({M_\star \over \mathrm{M_\odot}}\right)  - 2.61 \ ,
\end{split}
\end{equation}
\begin{equation}
\label{eq:02}
\begin{split}
& \log\left({r_{1/2, \mathrm{\, min}} \over \mathrm{kpc}}\right) = -0.06 \,\log\left({M_\star \over \mathrm{M_\odot}}\right)  + 0.33 \ ,
\end{split}
\end{equation}
for galaxies in $8.4 < \log(M_\star/\mathrm{M_\odot}) < 9.2$. Finally, we defined a conservative control sample of galaxies (hereafter,  ``\Normals'') as the galaxies within 25th - 75th percentiles of the Main branch. As the \CompactMBs\ have a higher median stellar mass than the other two samples, we selected subsamples of \Normals\ and  \CompactSBs\ to have the same stellar mass distribution as the  \CompactMBs{}, with median stellar mass $10^{8.75} \msun$ within two effective radii; that is, $10^{8.9}\msun$ over the full subhalo. Our selection is shown in Fig.~\ref{fig:sizevsmass_classes}.

This selection led to 1289 \Normal s,\, 131 \CompactMBs,\ and 157 \CompactSBs\  galaxies. We also split these samples (hereafter, ``size classes'') according to their location  within their group at $z=0$, between $z$=0 ``centrals'' and ``satellites''. We abusively use these terms to refer to the respective $z = 0$ galaxies and their main progenitors, although $z=0$ satellites were centrals at earlier times. We identify centrals as the subhalos whose index is contained in the list of central subhalos of all $z$=0 groups (obtained from {\tt GroupFirstSub}). We verify centrals that are not  backsplash  galaxies (those that previously traveled once or several times  through a group and are currently outside that group and are identified as the center of another  group).  For this, we verified that the group mass of the central  ({\tt Group\_M\_Crit200}) was not more than 1.5 dex greater at earlier times.  The fractions of centrals are  $54\%$, $37\%$, and $52\%$ among \CompactsSB, \CompactsMB, and \Normals, respectively. In other words, while roughly half the $z$=0 \CompactsSB{} and \Normals{} are centrals, almost two-thirds of the $z$=0 \CompactsMB{} are satellites.

\begin{table}[h]
\caption{Samples of \Compacts\ and \Normals}    
\label{tab:samples}     
\centering                                       
\tabcolsep=2.5pt
\begin{tabular}{lrrr}   
\hline\hline  
                                 &   \multicolumn{1}{c}{Central}  & \multicolumn{1}{c}{Satellite}  & \multicolumn{1}{c}{All} \\
\hline
\Normals:    &\\
Younger          & 0\ \ \    (0.0\%)  & 1 \ \ (0.1\%) &1\ \ \ \ \ (0.1\%) \\
Intermediate & 3\ \ \    (0.2\%)  & 0 \ \ (0.0\%) &3\ \ \ \ \ (0.2\%)  \\
Old              & 674     (52.3\%) &   611    (47.4\%) &1285\ \ \ (99.7\%) \\
\cline{1-4}
Total            &  677    (52.5\%) & 612     (47.5\%) &1289 (100.0\%) \\
\hline
\CompactsMB: & \\
Younger          &  0\ \ \   (0.0\%)   & 0\ \ \    (0.0\%)&0\ \ \  \ \  (0.0\%) \\
Intermediate &  0\ \ \   (0.0\%)   & 0\ \ \    (0.0\%) &0\ \ \ \ \   (0.0\%) \\
Old              &  48     (36.6\%) & 83         (63.4\%)&131 (100.0\%) \\
\cline{1-4}
Total            &  48     (36.6\%) & 83         (63.4\%) &131 (100.0\%) \\
\hline
\CompactsSB:    &\\
Younger          & 0\ \ \    (0.0\%)  & 14\ \ \,(9.0\%) &14\ \ \ \ \ (9.0\%) \\
Intermediate & 0\ \ \    (0.0\%)  & 1\ \ \,(0.6\%) &1\ \ \ \ \ (0.6\%)  \\
Old              & 85     (54.1\%) & 57\ (36.3\%) &142\ \ \ (90.4\%) \\
\cline{1-4}
Total            &  85    (54.1\%) & 72\ (45.9\%) &157 (100.0\%) \\
\hline
Bad-flag:    &\\
Younger          & 0\ \ \    (0.0\%)  & 130 (100.0\%) & 130 (100.0\%) \\
Intermediate & 0\ \ \    (0.0\%) &  0\ \ \    (0.0\%)  &  0\ \ \    (0.0\%)  \\
Old              & 0\ \ \    (0.0\%) &  0\ \ \    (0.0\%)   &  0\ \ \    (0.0\%)  \\
\cline{1-4}
Total            &  0\ \ \    (0.0\%) & 130 (100.0\%) &130 (100.0\%) \\
\hline
\end{tabular}

\parbox{\hsize}{}
\end{table}

 We also defined galaxies that were born later than $z = 1$ as the ``{young}'' population; galaxies that were born between $z=5$ and $z=1$ as the ``{intermediate'}' population; and galaxies that were born before  $z=5$ as the ``{old}'' population. Most galaxies are old: $90\%$, $100\%,$ and $99.7\%$ for \CompactsSB{}, \CompactsMB, and \Normal,\ respectively.  No \CompactsMB{} have an intermediate or younger age. Among the \Normals, only three have an intermediate age, while only one is younger. Finally, only one \CompactSB{} has an intermediate age, while $9\%$ are younger, all of which are satellites. As we have few galaxies from the intermediate-age population, we exclude these from our analysis.

Table~\ref{tab:samples} lists the numbers of galaxies in each size class and subsample
of our final sample (with percentages in parentheses). None of the central galaxies have a young age or a bad flag, and only 3 among 810 have intermediate ages. In the present work, we only consider  the central old galaxies. After the above selections, we therefore considered 674 \Normals, 48 \CompactsMB{}, and 85 \CompactsSB{}.

\subsection{Components of $z$=0 galaxies}

\label{sec:components}
\begin{figure}[ht]
  \centering
  \includegraphics[width=0.9\hsize]{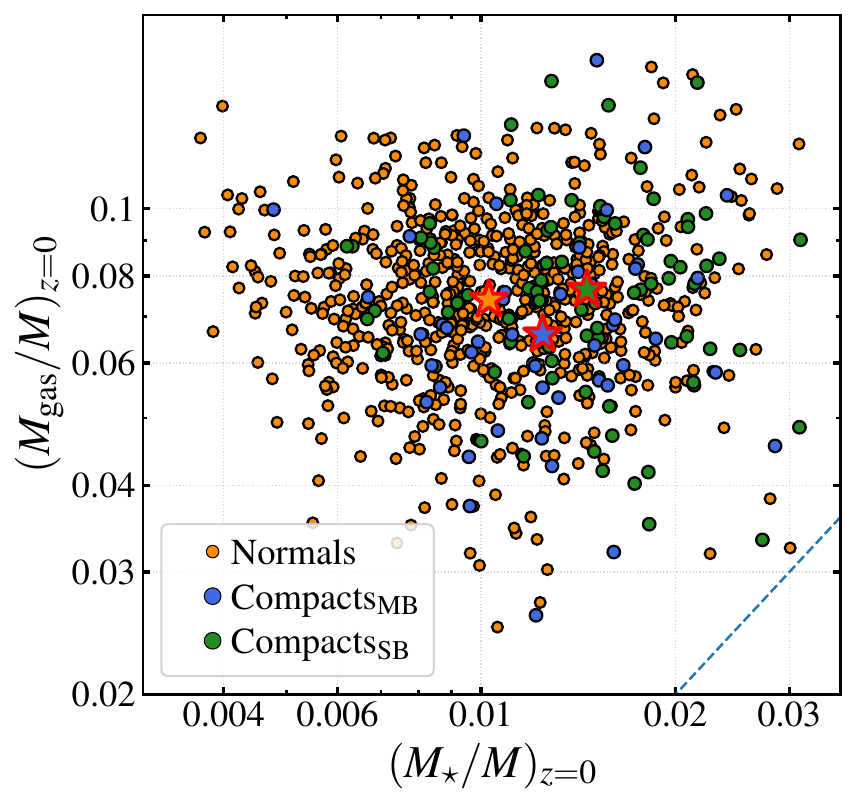} 
  \caption{Gas fraction versus stellar fraction  at $z = 0$. The masses of each component are summed over the full subhalo. The {orange}, {blue,} and {green circles} are the different size classes, respectively: \Normals,\, \CompactsMB\ , and \CompactsSB. The {blue dashed} line indicates equality. The {star symbols} indicate the median values for the gas and stellar mass fractions.
  }
\label{fig:GasStarDMFracs}
\end{figure}

A more global view of the relative importance of the three components (stars, gas, and DM) among central galaxies is given in  Table \ref{tab:DMGasFrac}, which shows the median $z$=0 DM, gas, and stellar fractions in each population. In Table~\ref{tab:DMGasFrac} and elsewhere in the present paper,  we verify the significance of different distributions using  5000 random shuffles of the members of the two samples. The central \CompactsSB{}, \CompactsMB,{} and \Normals{} are typically dominated by DM. \Normals{} and \CompactsMB{} have the same distribution for each component, while \CompactsSB{} have slightly more stars (median of 0.02) and less DM (median of 0.91), with statistically different distributions. Figure~\ref{fig:GasStarDMFracs} shows  the present--day fractions for the gas and stellar components, with no striking differences between our three size classes. 

\begin{table}[h]
\caption{Median $z$=0 DM, gas, and star fractions for central galaxies}                 
\label{tab:DMGasFrac}    
\begin{center}     
\tabcolsep=3pt
\begin{tabular}{lccc}   
\hline\hline  
                                 &   \multicolumn{1}{c}{Dark matter}  & \multicolumn{1}{c}{Gas}  & \multicolumn{1}{c}{Stars} \\
\hline
\Normals{} & 0.92   &   0.07   &  \textbf{0.01}  \\
\CompactsMB{} & 0.92   &    \textbf{0.07}   &   \textbf{0.01}     \\
\CompactsSB{} &  \textbf{0.91}   &  0.07   &   \textbf{0.02}   \\
\hline
\end{tabular}

\parbox{\hsize}{Notes: The bold values indicate that the distribution of a given component for one class is significantly different
  ($P<0.05$) from those of the other two classes, using 5000 random shuffles.
  }
\end{center}
\end{table}

\section{Differences in the histories of \Compact\ and \Normal\ galaxies}
\label{sec:evol}

We explored how galaxies become compact by studying the backwards  evolution (hereafter, ``history'') of specific parameters of the most massive progenitors (hereafter, ``main progenitors'') of the $z$=0 galaxies, separately for the \Compacts\ and \Normals. We  follow the main progenitors using the Main Progenitor branch of the {\sc SubLink} merger trees from the TNG database.

We first studied the evolution of  the median (hereafter, ``median evolution'') of specific galaxy parameters to understand the different histories by tracking the histories of each of the 1289 galaxies that end up as centrals (always being centrals throughout their history) ---that is, 674 \Normal, 48 \CompactMBs, and 85 \CompactSBs{}--- and taking the median at each snapshot containing at least five galaxies (their main progenitors). We used bootstraps to estimate the uncertainties on these medians.

\subsection{Size and mass evolution}

\begin{figure}[htpb!]
    \center
    \includegraphics[width=\hsize]{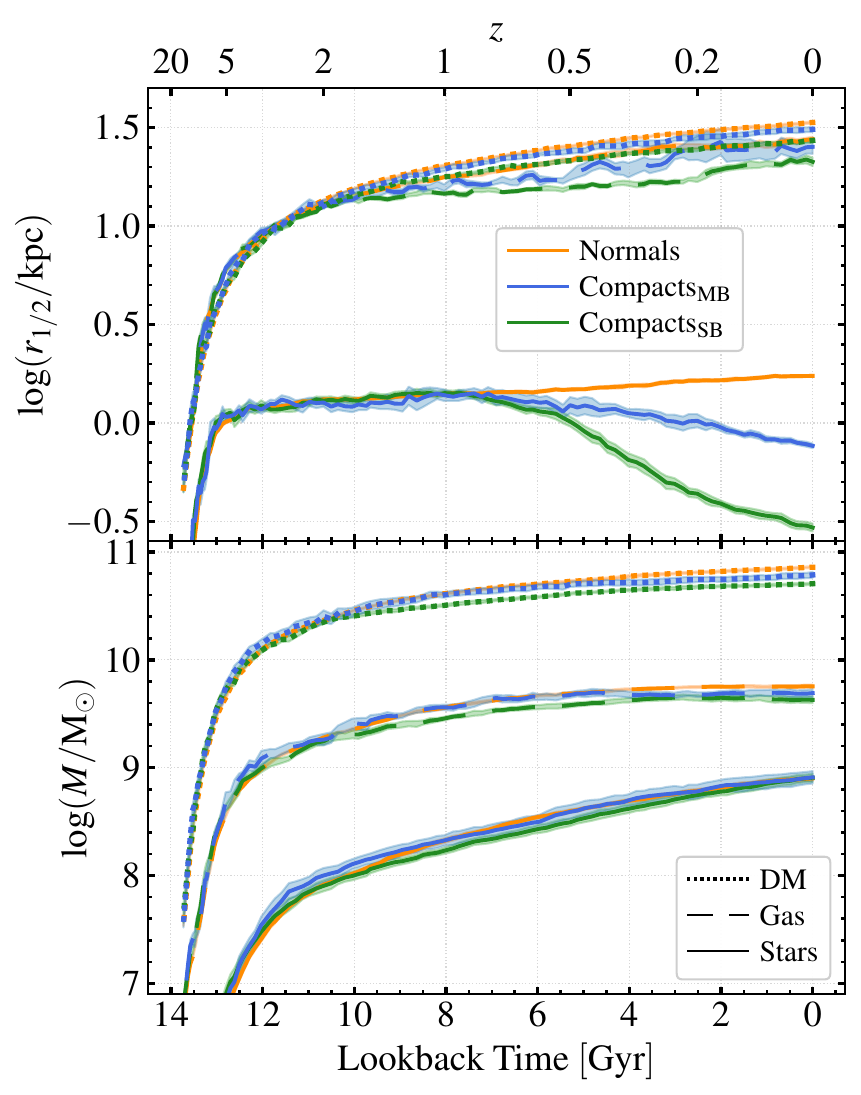}
    \caption{Median evolution of the half-mass radii (\textbf{top}) and total masses (\textbf{bottom}) of the \Normals\ ({orange}), \CompactsMB{} ({blue}), and \CompactsSB{} ({green}) for galaxies that end up as centrals at $z=0$. The {solid}, {dashed}, and {dotted lines} represent the stellar, gas, and DM components, respectively,  and all quantities are measured over the entire subhalos. The {lines} are the medians, and are only shown when we have at least five galaxies at that epoch;  the {shaded region} shows the uncertainty on the median, estimated using bootstraps.
}
  \label{fig:sizemassevol}
\end{figure}

Figure~\ref{fig:sizemassevol} shows  the median evolution of the sizes and  masses  of the stellar, gas, and DM components measured over the entire subhalo. We do not use the popular TNG stellar mass within twice the stellar half-mass radius, which is clearly size dependent and would tend to show decreasing stellar mass evolution for the \Compacts{}.

The solid lines and associated shaded regions of the top panels of Fig.~\ref{fig:sizemassevol} show the median evolution of the stellar half-mass radius.  One should note that  galaxies that are centrals at $z=0$ were not satellites at earlier epochs, as we removed the $z$=0 backsplash centrals.

The top panel of Fig.~\ref{fig:sizemassevol} indicates that the stellar size of \Normals\ grows at all times. In contrast, the \CompactsMB{} and \CompactsSB{} both start shrinking at $z=0.8$. This behavior appears robust, as it is also seen when we split our samples into five bins of final stellar mass. We then verified that the sizes of the stellar components of \Compacts{} and \Normals{} are very different  at $z = 0$, as expected from our selection (with \CompactsMB{} in between the \Normals\ and the \CompactsSB{}). But at $z\sim0.8$, the median stellar half-mass radii of the three size classes are similar.
This decrease in size (hereafter, compaction) is roughly exponential in time (linear in the figure) between $z=0.5$ (5.2 Gyr ago) and $z=0.3$ (3.5 Gyr ago). The size reduction is a factor 1.9 for \CompactsSB{} in this 1.7 Gyr time interval, and is a factor 1.1 for \CompactMB\ centrals. In summary, on average, \Compact{} galaxies were not compact to begin with, but evolved to become compact.

Interestingly, as seen in the top panel of Fig.~\ref{fig:sizemassevol}, while \CompactsSB{} show a huge decrease in  stellar half-mass radius, the corresponding DM and gas half-mass radii of \Compacts\ (upper panel) evolve relatively similarly to those of the \Normals{}: the $z$=0 DM sizes are 5\% and 20\% lower for \CompactsMB\ and \CompactsSB, respectively, while the $z$=0 gas sizes are 8\% and 30\% lower, respectively.   

The lower panels of Fig.~\ref{fig:sizemassevol} show the median total mass evolution of the different components of the different size classes. The total mass corresponds to the mass within the entire subhalo, and not only within 2 stellar half-mass radii. One sees that the sample has very little effect on the typical mass evolution of all three components of the galaxies.  As we selected our sample using the stellar mass within $2\,r_{1/2}$, we can deduce that the outer masses of \CompactsSB{} are slightly depleted.

Combining the size and mass evolution in \Compacts\ that end up as centrals, the compaction of their stellar components occurs independently of the gas and DM components, of the \Compact\ size class, and of the mass evolution of the DM, gas, and even stellar components.

\subsection{Evolution of specific star formation rates}
\label{subsec:ssfr}

\begin{figure}[htpb!]
    \center
    \includegraphics[width=\hsize]{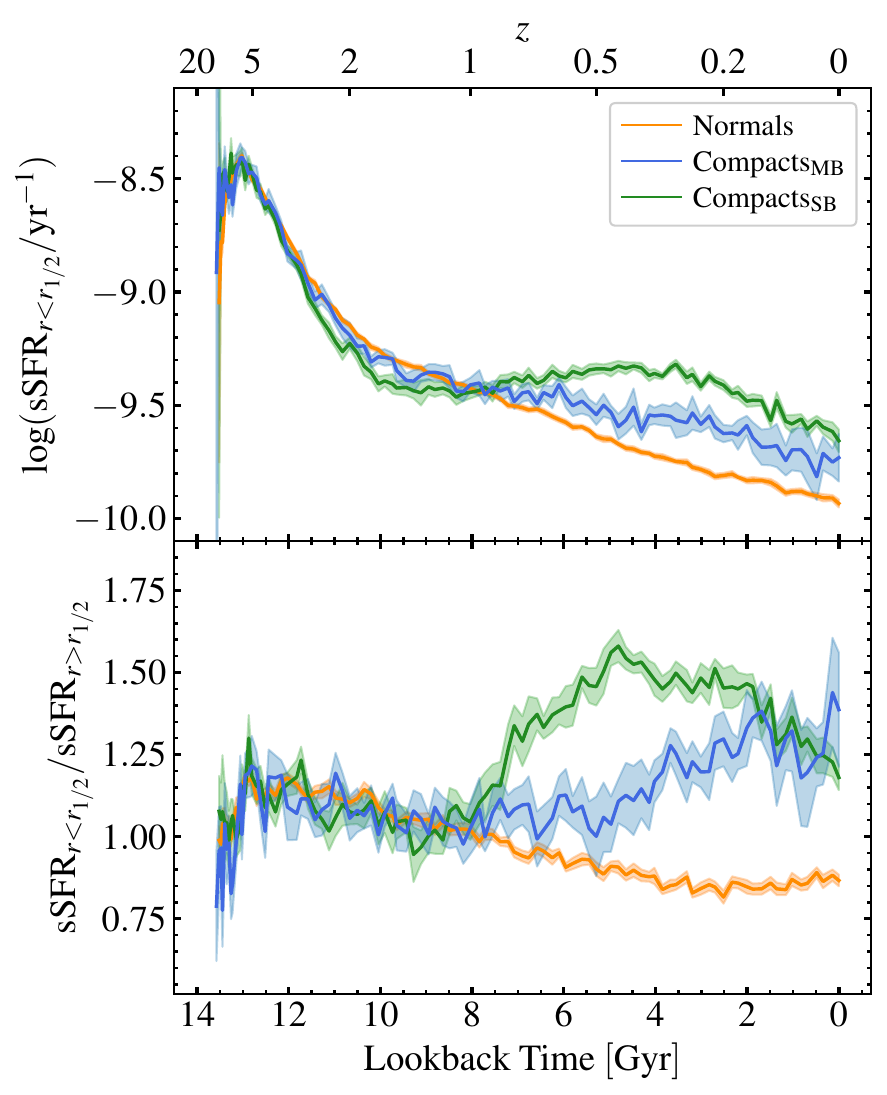}
    \caption{Same as  Fig.~\ref{fig:sizemassevol}, but for the median evolution of the sSFR in the inner region ($r < r_{1/2}$)  (\textbf{top}) and the ratio between the sSFRs in the inner and outer regions ($r > r_{1/2}$) (\textbf{bottom}).
  }
  \label{fig:ssfrevol}
\end{figure}

Some of the differences in the evolution of the stellar mass component may be
related to differences in the evolution of the star formation efficiency (SFE). Here, we consider the {specific star formation rate} (sSFR; star formation rate over stellar mass) as a proxy for SFE. Figure \ref{fig:ssfrevol} compares the histories of the sSFR of \Compacts\ and \Normals, showing the sSFR in the inner region ($r < r_{1/2}$) and the ratio between that sSFR and the sSFR in the outer region ($r > r_{1/2}$).  We estimate the sSFRs using the stellar masses and SFRs provided in the TNG database in the full subhalo and within $r_{1/2}$.

Figure \ref{fig:ssfrevol} shows that the main progenitors of the $z$=0 \Normals\ have a continuous decrease in their inner sSFR (i.e., ``{quenching}'' of star formation) with cosmic time. There are  small gradients in sSFR at $z \sim 1$ for this population: the ratio between the inner and outer sSFR is close to unity. The inner star formation starts to become lower than in the outer region after $z \sim 0.9$. The \CompactsMB\ also see a continuous decrease in their inner sSFR, but not as strongly as do the \Normals{}.
 
In contrast, the inner sSFR in \CompactsSB\ is almost constant between $z \sim 1.7$ and $\sim 0.45$. This is related to concentrated star formation, which is verified by the rapid increase in the ratio between the inner and outer sSFR. \CompactsMB{} also have a slow increase in this ratio, which suggests that these galaxies also have concentrated star formation, but not as strong as in \CompactsSB{} and also without the almost constant inner sSFR. We return to these radial-dependent sSFRs in \Compacts\ in Sect.~\ref{sec:profiles}.

\subsection{Merger history}
\label{subsec:merger}
\begin{figure}[htpb!]
    \center
    \includegraphics[width=\hsize]{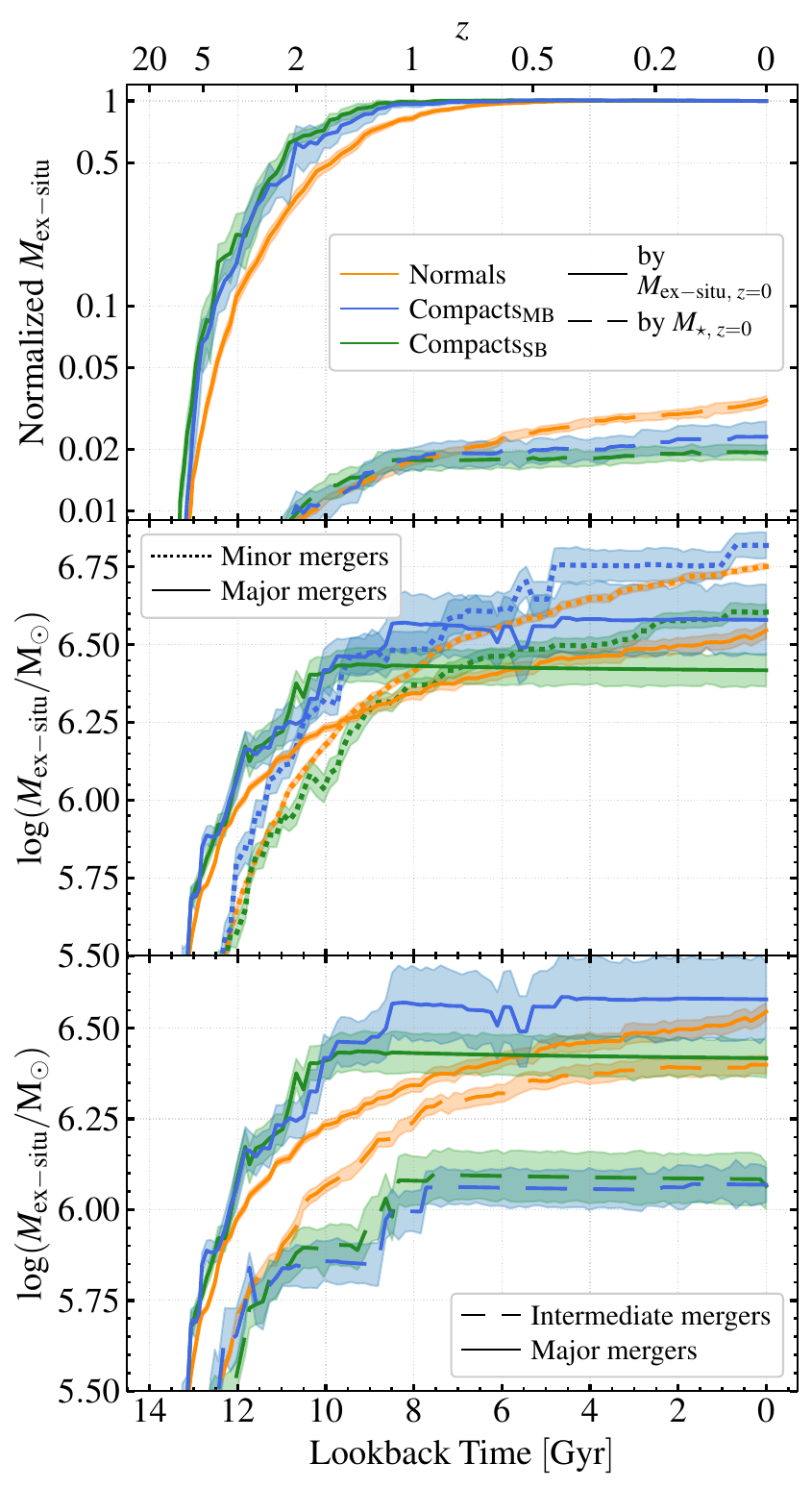}
    \caption{Same as Fig.~\ref{fig:sizemassevol}, but for the median evolution of the ex situ stellar mass. {\bf Top}: Ex situ mass normalized by the final ex situ mass ({solid}) or   final  (ex situ + in situ) stellar mass (i.e., ex situ stellar mass fraction, {dashed}). All masses are   measured over the entire subhalo. {\bf Center} and {\bf bottom}: Ex-situ stellar mass from major mergers (stellar mass ratio $> 1/4$, \textit{solid} lines), intermediate  mergers (stellar mass ratio  between $1/4$ and $1/10$, {dashed} lines), and minor mergers (stellar mass ratio below $1/10$,  {dotted} lines).
 }
  \label{fig:mergeevol}
\end{figure}

The star formation in galaxies can be within the main progenitor ({in situ}) or within external galaxies that later merged with the main progenitor ({ex situ}). The top panel of Fig.~\ref{fig:mergeevol} displays the median evolution of the ex situ stellar mass normalized by the $z$=0 ex situ stellar mass (solid) and by the $z$=0 total stellar mass (dashed); while the bottom panel show the stellar mass from major mergers and other mergers. The data were obtained using the {\sf Stellar Assembly TNG} supplementary data catalog provided by \cite{RodriguezGomez2016MNRAS, Rodriguez2017MNRAS}.

The dashed lines in the top panel of Fig.~\ref{fig:mergeevol} indicate that the ex situ fractions of stellar mass for all three size classes are typically small ($<4\%$) at all times. Furthermore, while the ex situ stellar mass fraction of \Normals\ is continuously increasing,  those of the two \Compacts{} size classes decouple from that of the \Normals{} at $z = 0.8$, quickly reaching their maxima.  By $z\,=\,0$, the ex situ fraction is 2\% and 2.5\% for \CompactsSB\ and \CompactsMB,\ respectively, compared to 3.5\% for \Normals. The narrow uncertainties (shaded regions) indicate that this slower ex situ growth of \Compacts\ is highly significant.
 
The solid lines of the top panel of Fig.~\ref{fig:mergeevol} indicate that \CompactsSB\ and \CompactsMB\ tend to collect half their ex situ mass roughly 1.4 and 1.0 Gyr earlier than \Normals, respectively, that is, by $z\,=\,2.3$ (11.0 Gyr ago) and $z\,=\,2.0$ (10.6 Gyr ago), respectively, instead of by $z\,=\,1.5$ (9.6 Gyr ago) for \Normals. Given the lower contribution from ex situ material, \Compacts{} rely more on in situ star formation to grow in stellar mass  compared to \Normals. In other words, mergers play a smaller role in the evolution of galaxies that end up \Compact.

The second and third panels of Fig.~\ref{fig:mergeevol} show the contributions of mergers of different mass ratios to the ex situ mass: ``{major}'' (1/4 to 1), ``{intermediate}'' (1/10 to 1/4), and ``{minor}'' mergers (less than 1/10).\footnote{Our merger nomenclature differs  from that of the TNG database, which denotes ``minor'' mergers as stellar mass ratios from 1/4 to 1/10 (what we call intermediate).} At $z = 0$, \CompactsSB{} typically have $\sim$20\% (0.1 dex) lower stellar mass from major mergers (green solid) than \Normals\ and \CompactsMB{} (orange and blue solid lines). Also, \Normals\ have a continuous growth of mass by major mergers, whereas \Compacts\ typically reach a maximum mass from major mergers at $z\sim 1.4$ (\CompactsSB) or 1.2 (\CompactsMB). Moreover, before $z=2$, the growth of stellar mass from major mergers is   faster in \Compacts\ than in \Normals{}. The bottom panel of Fig.~\ref{fig:mergeevol} shows that intermediate mergers are even less important for the \Compacts\ than for the \Normals: they provide roughly three times less ex situ stellar mass to the \Compacts{} than to the \Normals.

The {\sf Stellar Assembly TNG} also provides the flyby contribution in ex situ stellar mass. The flybys contribute less than 1\% of the ex situ stellar mass for the three size classes. Indeed,  the median mass provided by flybys in \CompactsSB{} is zero. In summary, the stellar build up of \Compacts, especially \CompactsSB{}, is less driven by major and intermediate mergers, while flybys are negligible for all three classes. 

 \begin{figure}[htpb!]
    \centering
\includegraphics[width=1 \hsize]{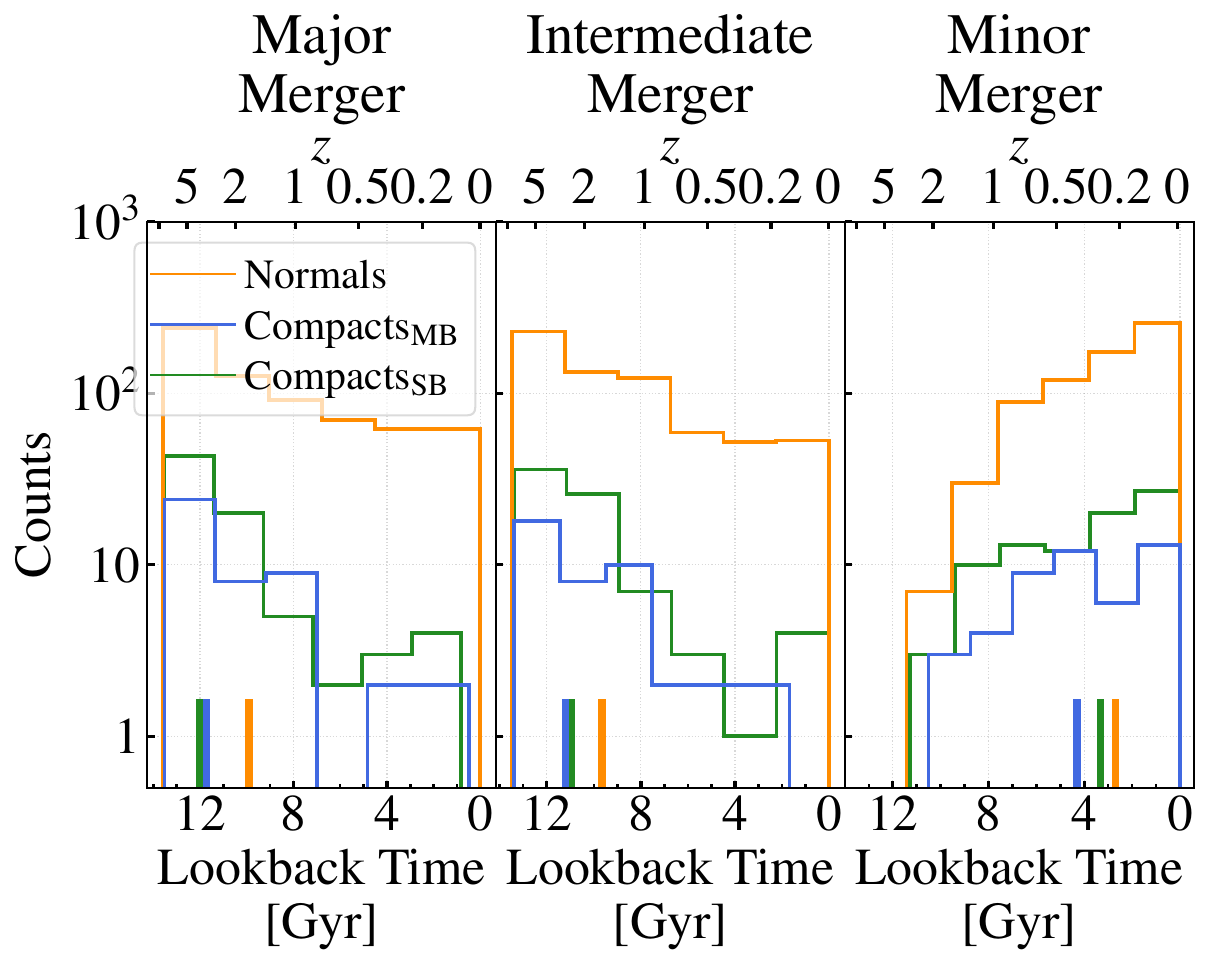}
    \caption{ Distributions of the epochs of the last mergers.The {\bf left}, {\bf middle}, and {\bf right} panels show the major (stellar mass ratio $M_2/M_1>1/4$), intermediate ($1/10 < M_2/M_1<1/4$), and minor ($M_2/M_1<1/10$) mergers, respectively.  
    The {vertical segments} indicate the median of each distribution. The colors are the same as in Fig.~\ref{fig:sizemassevol}.
 }
  \label{fig:histMergers}
\end{figure}

Figure \ref{fig:histMergers} compares the merger histories of \Compacts\ and \Normals, showing the epochs of the last major merger, intermediate merger, and minor merger. The last mergers of the three mass ratio classes occur earlier in \CompactsMB{} than in \Normals: $z_{\rm median} = 2.74$, 2.58, and 0.38 in \CompactsMB{} versus 1.6, 1.49, and 0.21 in \Normals, for major, intermediate, and minor mergers, respectively. \CompactsSB{} lie in between \CompactsMB{} and \Normals{} for intermediate and minor mergers, while these galaxies have earlier major mergers, with median redshifts of 3.28, 2.21, and 0.27 for the last major, intermediate,  and minor merger, respectively.

Table~\ref{tab:LastMerger} shows the significance of the differences between the medians of the redshift of the last merger.  The epoch of the last major merger of  \CompactsSB{}  is statistically identical to that of  \CompactsMB{}  ($P = 0.29$), while the last major merger occurs significantly later  in \Normals{} ($P = 0.0004$ and $0.01$, respectively). Similarly, the last intermediate merger occurs significantly earlier in  \CompactsSB{} and \CompactsMB{} than it does in \Normals{} ($P = 0.01$ and $0.01,$ respectively). In contrast, while the last minor merger occurs significantly earlier in \CompactsMB{} than it does in \Normals, this is not the case for \CompactsSB{}. Disregarding the minor mergers, \Compacts\ thus have more time to evolve without being appreciably perturbed by mergers, which explains why their ex situ stellar mass is  smaller than in \Normals{}. 

\begin{table}[h]
     \caption{Median redshift of last merger}
    \label{tab:LastMerger}
    \begin{center}
    \begin{tabular}{lccc}
    \hline
    \hline
   Merger class (mass ratio)
    & \Normals & \multicolumn{2}{c}{\Compacts}\\
    \cline{3-4}
    & & (MB) & (SB) \\
    \hline
        major  ($>1/4$) &  {\bf 1.60} & 2.74 & 3.28  \\
     intermediate  ($1/10$ to $1/4$) & {\bf 1.49} & 2.58 & 2.21 \\ 
     minor  ($<1/10$) & 0.21 & {\bf 0.38} & 0.27 \\
      \hline
    \end{tabular} 
   \end{center}
\end{table}

\subsection{Evolution of the environment}
\label{subsec:environment}

\begin{figure}[htpb!]
    \centering
\includegraphics[width=\hsize]{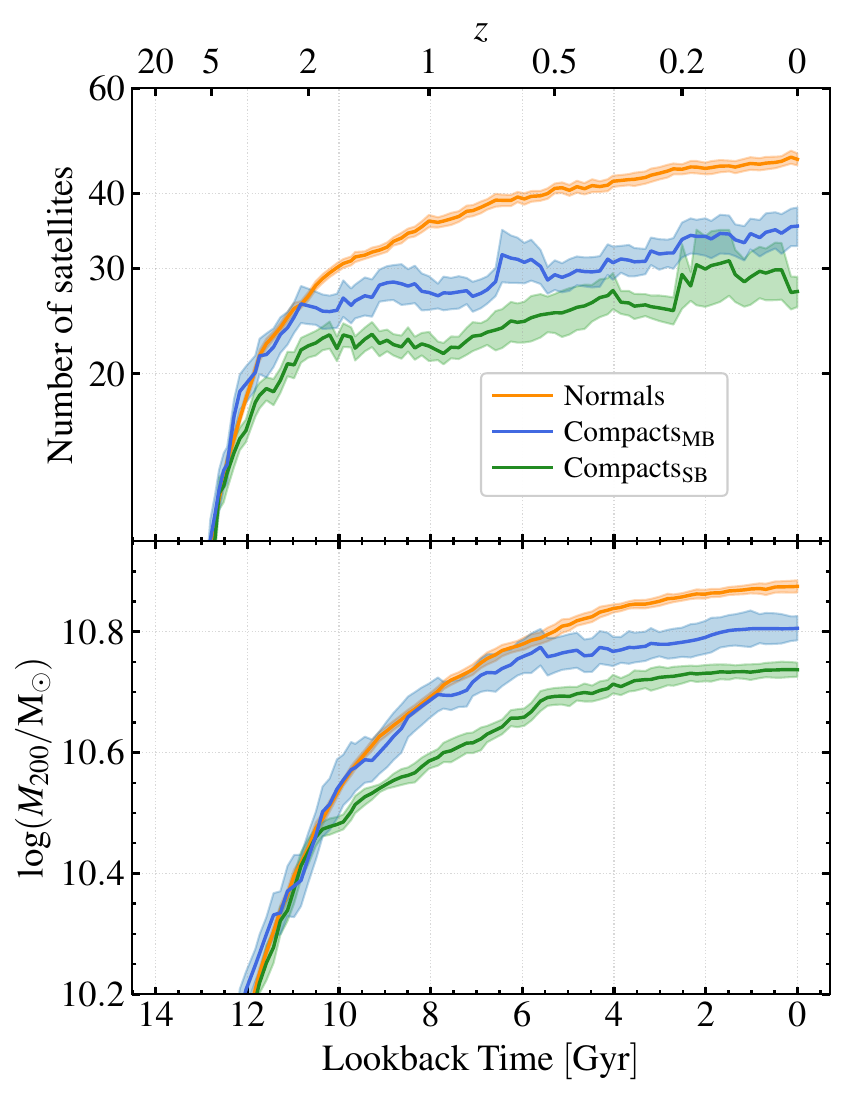}
    \caption{Same as Fig.~\ref{fig:sizemassevol}, but for the median evolution of the  satellite number (\textbf{top}) and halo mass $M_{200}$ (\textbf{bottom}) for central galaxies. }
  \label{fig:envtevol}
\end{figure}

The lower ex situ mass fraction and fewer (major) mergers of \Compacts{} relative to \Normals\ of the same $z$=0 stellar mass suggest that \Compacts\ live in lower density regions.  The top panel of Fig.~\ref{fig:envtevol} shows that  central \Compacts{} have lower numbers of satellites\footnote{We do not impose a threshold in mass on the satellites around our centrals.} than \Normals\ from $z=0.5$ onward: by 25\% (\CompactsMB) and 80\% (\CompactsSB). The lower panel indicates that central \Compacts\  live in  slightly lower mass halos than \Normals. At $z=0$, the halos of \Compacts\ are 25\% (\CompactsMB) to 60\% (\CompactsSB) lower in mass than those of \Normals.   The evolution of the number of satellites of \Compacts\  departs from that of \Normals\ at $z\sim2$ (\CompactsMB) and $z\sim5$ (\CompactsSB), with a plateau or even a decreasing number until $z\sim0.3$. Similarly, the evolution of the halo mass of \Compacts\  departs from that of \Normals\ at $z\sim1$ (\CompactsMB) and $z\sim2$ (\CompactsSB), with continuously increasing halo mass for all three size classes. In summary, \Compact\ centrals, \CompactsMB,{} and especially \CompactsSB live in slightly lower mass groups and  have fewer satellites.

\subsection{Morphological evolution}

\begin{figure}[htpb!]
    \centering
    \includegraphics[width=1\hsize]{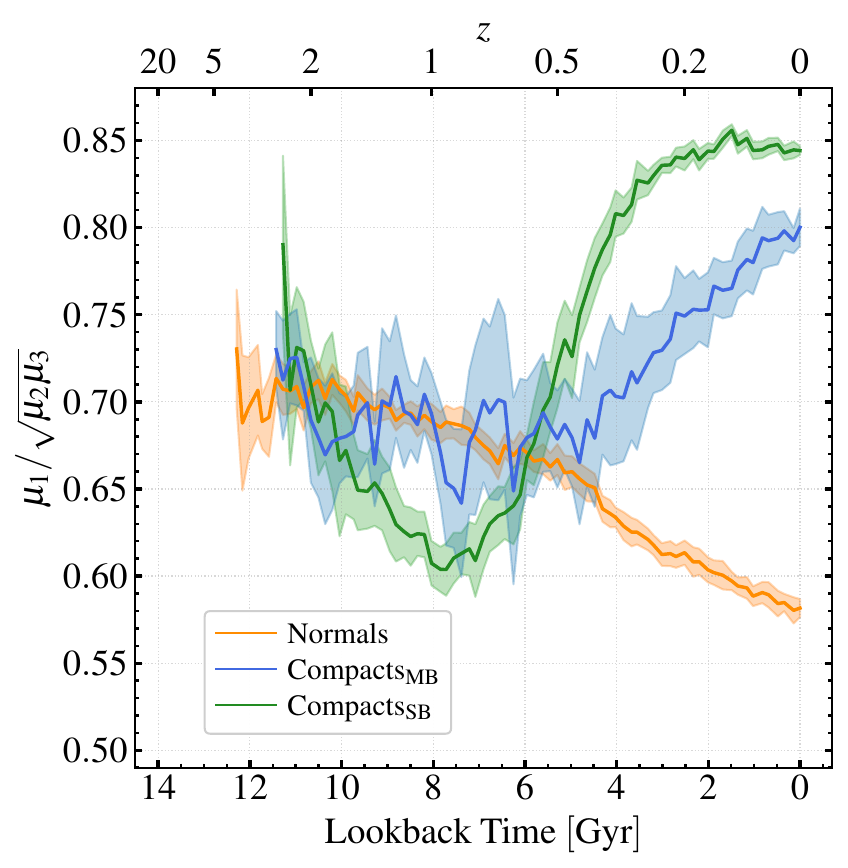}
    \caption{Same as Fig.~\ref{fig:sizemassevol}, but for the median evolution of the sphericity of the stellar distribution for all galaxies. The high (respectively low) values  indicate spherical (flat) galaxies.
    }
  \label{fig:shapeevol}
\end{figure}

We follow the evolution of the flattening of the stellar distribution using the {\sf Stellar Circularities,  Angular Momenta, Axis Ratios} TNG supplementary data catalog \citep{Genel2015ApJ}. This catalog provides the eigenvalues, $\mu_i$, of the mass tensor of the stellar mass within $2\,r_{1/2}$, with $\mu_1 < \mu_2 < \mu_3$. We define a {sphericity} parameter, $\mu_1 / \sqrt{\mu_2 \, \mu_3}$, where low values indicate a flattened galaxy \citep[as discussed by][]{Genel2015ApJ}.

Figure \ref{fig:shapeevol} displays the median evolution  of the sphericity parameter  for the \Compacts\ and \Normals{}. The main progenitors of the \Normal\ centrals typically become progressively flatter in time. Relative to the \Normal\ centrals, the \CompactMB\ centrals follow the same trend in shape until $z\sim0.6$, when they become rapidly more spherical. In contrast, the \CompactSB\ centrals flatten more quickly than the other two size classes, until $z\sim 0.8$, when they very rapidly become more spherical until 2 Gyr lookback time and remain so until the present day.

\subsection{Individual histories}

\begin{figure*}[htpb!]
    \center
   \includegraphics[width=\hsize]{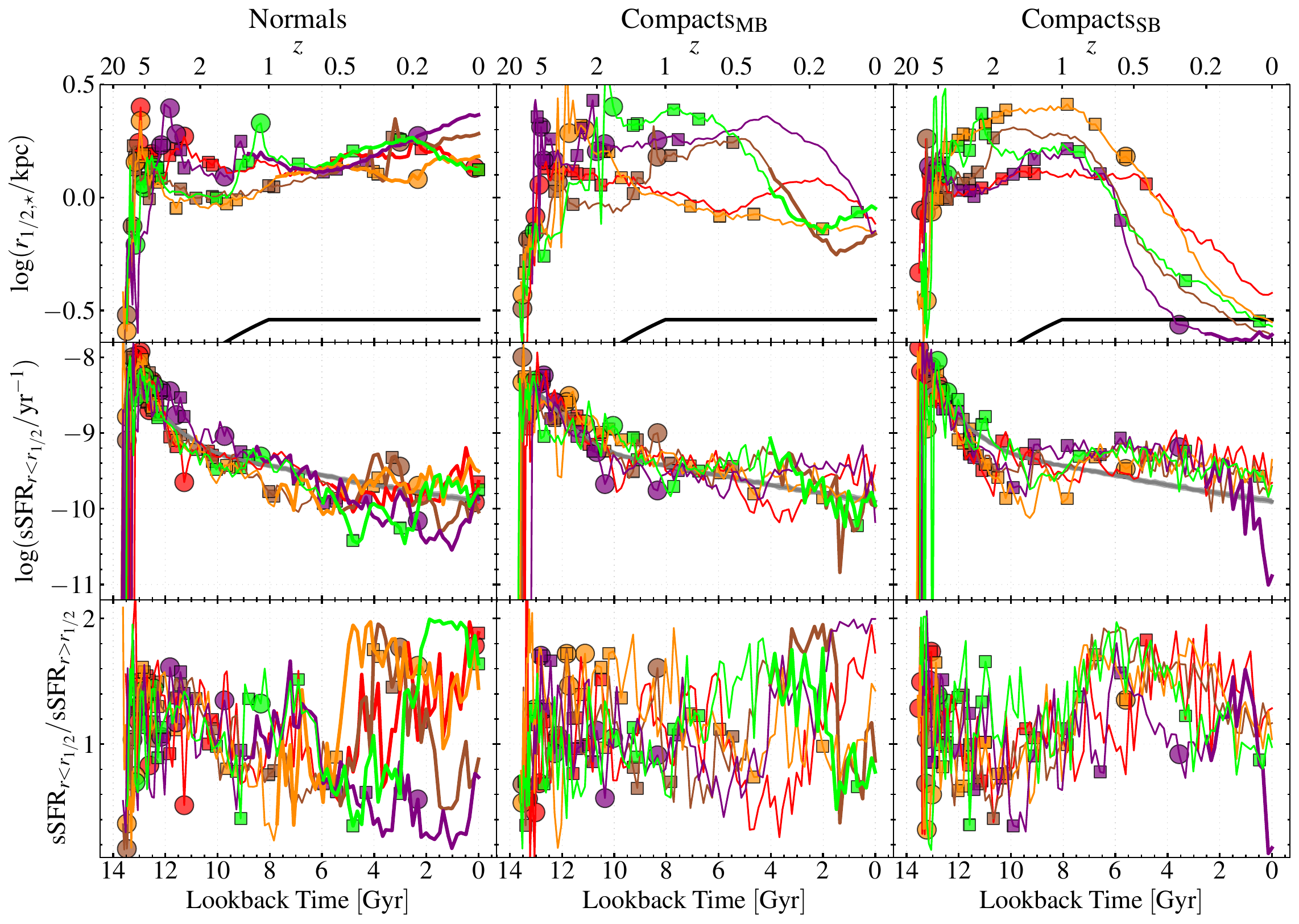}
    \caption{Individual histories of the half-mass radius (\textbf{top}), inner sSFR (\textbf{second row}), and the ratio between the sSFRs in the inner and outer regions (\textbf{bottom}) for five randomly selected \Normals{} (\textbf{left}), \CompactsMB{} (\textbf{center}), and  \CompactsSB{} (\textbf{right}). Each line corresponds to the evolution of a single galaxy. The {symbols} indicate major mergers   (\textit{large circles}, with stellar mass ratio $> 1/4$) or intermediate mergers   (\textit{intermediate-size squares}, with stellar mass ratio between $1/10$ and $1/4$). The minor mergers are not shown for clarity. {Thicker lines} indicate when the galaxy hosts a central BH. The {thick black} lines show the evolution of the  softening radius, while the {gray shaded regions} show the $\pm$1$\,\sigma$ region for the sSFR median evolution of all galaxies in the stellar mass range: $8.4 < \lmsun < 9.2$. 
      }
  \label{fig:indiv_compact}
\end{figure*}

While the median history of a given physical galaxy  parameter allows us to view the typical evolution of that parameter  for \Compacts\ and  for \Normals, each galaxy has its own evolution. We followed the histories of a few randomly selected individual galaxies to better understand the rapidity of the   decrease in size of the stellar component and its relation to the histories of other galaxy parameters.

Figure \ref{fig:indiv_compact} displays the individual histories of the half-mass radius and sSFR  for five \CompactsSB{}, \CompactsMB,{}  and  \Normal\ galaxies. The top panel of Fig.~\ref{fig:indiv_compact} shows that most \CompactsSB{} begin shrinking at a redshift of close to $z=0.8$ (except for the red one). The compaction of \CompactsMB\  occurs at different epochs: both with earlier (before $z \sim 1$) and late (after $z \sim 1$)  compaction. In contrast, the compaction of the different individual \CompactsSB\ appears  more synchronized in time.  On the other hand, all \Normals\  have relatively constant size  histories after $z\sim1$, with two cases of  centrals that somewhat shrink  after $z \sim 0.2$. All the compaction episodes starting later than 10 Gyr ago last at least 2 Gyr. Nearly all \CompactsSB{} decrease in stellar size over the last 8 Gyr. Thus, compaction is not a sudden process,  which is expected from Fig.~\ref{fig:sizemassevol}.

The compaction of \Normals\ and \CompactsMB\ is not affected by the softening of the gravitational interactions between stars and DM particles, as the softening scale (black curve) is always well below the stellar half-mass radius of the galaxies. On the other hand, the sizes of some of the \CompactsSB\ reach the softening scale at $z \sim 0.3$. This may affect subsequent evolution, as below this scale the calculated parameters are not necessarily reliable. We discuss this further in Sect.~\ref{sec:resolution}. The \CompactsSB\ continue to shrink until $z=0$, except one central (purple)  that suffers a major merger and stops shrinking. Many of the \CompactsMB\ also keep shrinking until $z=0$. Two \CompactsMB\ centrals (brown and green) stop shrinking when a BH is present (thicker curve).

In general, Fig.~\ref{fig:indiv_compact} confirms the median trend of Fig.~\ref{fig:histMergers} that major and intermediate mergers are rare in \Compacts\  after $z = 1$ (only two \Compacts\ out of 10 suffer a major merger). In contrast, mergers are more frequent in \Normals\ after $z=1$ (four out of five suffer a major merger). These results are in agreement with Fig.~\ref{fig:histMergers} and Table~\ref{tab:LastMerger}, which show that \Compacts{} stop undergoing mergers earlier than \Normals{}.

While AGN activity (thicker lines in Fig.~\ref{fig:indiv_compact}) is very rare in our ten individual \Compacts\ (only one \CompactsMB{} case at $z < 0.5$ and one \CompactsSB{} case at $z < 0.2$), it is more frequent in the \Normals, in particular at recent times ($z < 0.7$). In the top panel of Fig.~\ref{fig:indiv_compact},  we note that the size of \Normals\ increases roughly 1 Gyr after the onset of AGN activity. However, since the AGN activity is usually preceded by  a major or intermediate merger, it is not clear if this size increase is  driven by the merger or by the subsequent AGN activity. We return to this issue in Sect.~\ref{sec:bhagn}.

 \begin{figure*}[htpb!]
    \center
    \includegraphics[width=\hsize]{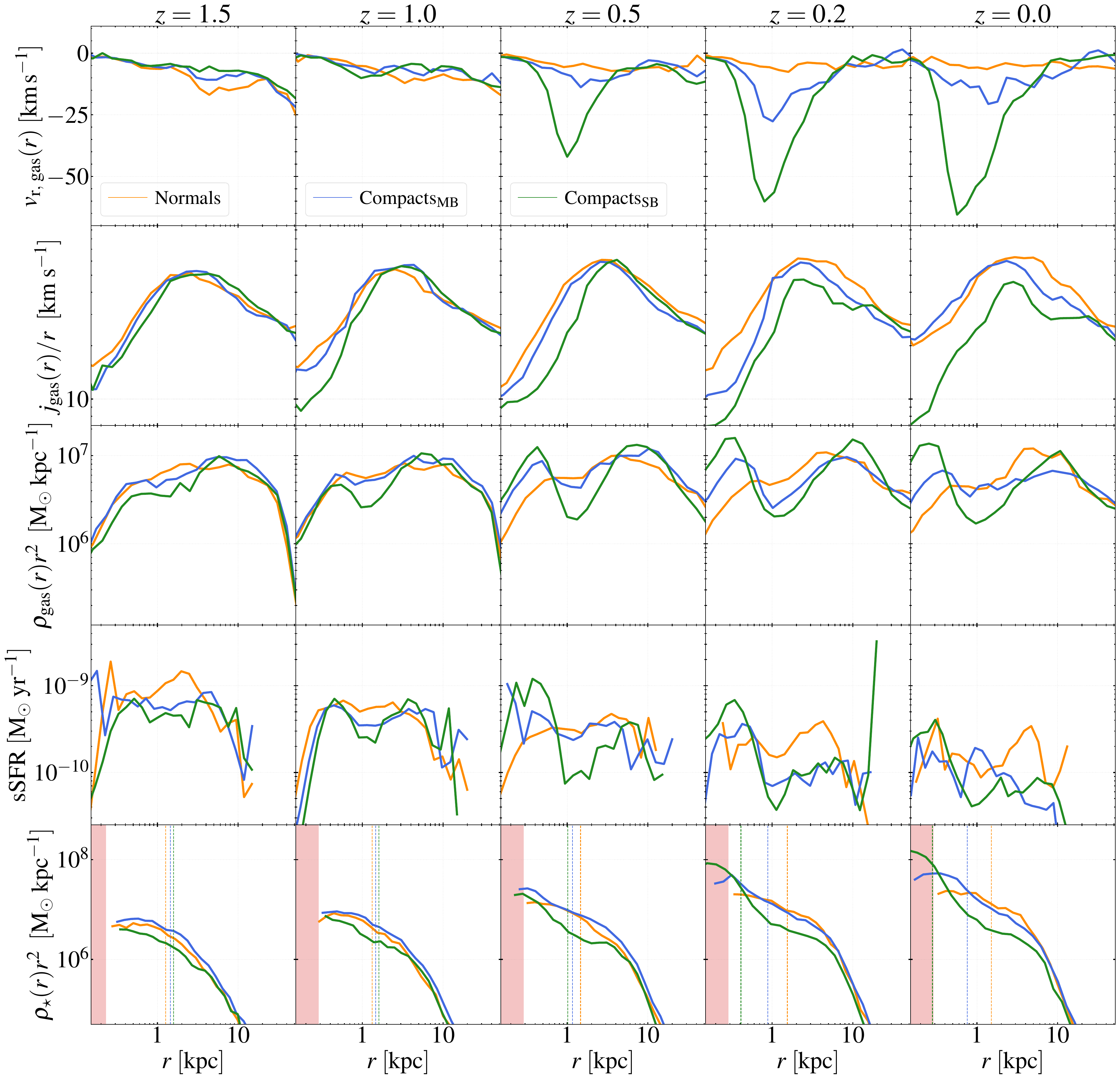}
    \caption{Median radial profiles of physical quantities of 30 random galaxies (10 random galaxies for each size class) that end up as centrals at different epochs: radial velocity of the gas  (\textbf{first} row),  gas specific angular momentum normalized by the radius (\textbf{second} row), gas density times $r^2$ (\textbf{third} row), sSFR  (\textbf{fourth} row), and stellar density times $r^2$ (\textbf{fifth} row). The $r$ values have been converted to physical kiloparsecs. The colors are the same as in Fig.~\ref{fig:sizemassevol}.  The {pink shaded regions} indicate the redshift-dependent, poorly resolved radii and the {dashed vertical} lines indicate the  redshift-dependent stellar half-mass radius.}
  \label{fig:FirstProfile}
\end{figure*}

As in Fig.~\ref{fig:ssfrevol}, the middle panels of Fig.~\ref{fig:indiv_compact} show that \Compacts{} (especially \CompactsSB{}) are able to maintain star formation in their inner region after $z = 1$ at an almost constant level and above the median sSFR evolution for all galaxies in the stellar mass range. We also note that epochs of late gradual compaction of individual \Compacts{} tend to occur during times when the inner sSFR is enhanced. For example, the green \CompactMB{} shrinks between lookback times of 6 and 2.5 Gyr, which corresponds to a higher inner sSFR compared to the median sSFR evolution of all galaxies in our adopted range of stellar masses. This panel is also interesting as it shows both rapid fluctuations in inner sSFR, especially for \Compacts{}, and  also  that individual galaxies can move above and below the median trend by typically 0.4 dex, and sometimes downwards by 0.8 dex. 

The bottom panels of Fig.~\ref{fig:indiv_compact} show the  evolution of the ratio between the sSFR in the inner region and that in the outer regions of the galaxy (in terms of $r_{1/2}$). \Compacts{} (especially \CompactsSB{}) have concentrated star formation after $z=1$ (ratio above 1). In particular, the periods of gradual compaction of \Compacts\ are even better matched to the ratio of inner to outer sSFR than to the inner sSFR itself (middle panel). For example, the brown \CompactSB{} rapidly shrinks between lookback times of 7 to 5 Gyr, when the inner sSFR is 1.8 times the outer one; and continues to shrink more gradually in the last 5 Gyr, when the inner sSFR is still 1.4 times the outer sSFR.

\subsection{Radial profiles}
\label{sec:profiles}

 Figure \ref{fig:FirstProfile} shows the median radial profiles at different redshifts  for the radial velocity of the gas, gas specific angular momentum normalized by the radius, gas density,  sSFR, and stellar density.  
 
 \CompactsSB{} have an intense gas infall compared to the other  galaxies starting at $z \sim 0.5$. This infall seems to be related to the lower angular momentum in the inner regions of the \Compacts{} (as shown in the second row of Fig.~\ref{fig:FirstProfile}): at $z \sim 0.5$, \CompactsSB{} have a specific angular momentum at $r = 1$ kpc that is only half of that of \Normals\ and \CompactsMB{}. \CompactsMB{} also show lower angular momentum, but only at $z \sim 0.2$ for $r \la 0.7$ kpc. The higher angular momentum of \Normals\  prevents the gas from efficiently infalling.

 As a result of these trends in gas infall, the gas density profiles (third row of Fig.~\ref{fig:FirstProfile}) of \CompactsSB{} and \CompactsMB{} display a relative under-density at $r \sim 1$ kpc for $z=0.2$ (about three times lower than the \Normals), while these galaxies have instead denser gas than \Normals\ in the inner regions (below $r \sim 0.7$ kpc). 

The gas kinematics should have a strong influence on the SFR. The fourth row of Fig.~\ref{fig:FirstProfile} shows the evolution of the radial profile of sSFR. While \Normals\ display only small gradients in sSFR at all times, \CompactsSB{} have a double-peaked sSFR profile at redshifts of unity and lower, with a deficiency in the intermediate region ($r$$\sim$1.5 kpc) at $z=1$ and 0.5 (by respective factors of three and over ten) relative to peak sSFR efficiencies at $r$$\sim$0.5 and $\sim$4 kpc.  \CompactsMB{} also display  this relation, but only after $z \sim 0.2$. The sSFR gradients agree with those seen in Fig.~\ref{fig:ssfrevol}. This shows that the compaction of \CompactsSB{} is related to intense star formation in the core. Central \CompactsMB{} have the same evolution, albeit not as strong as for \CompactsSB{}. This suggests that \CompactsSB{} are extreme cases of central \CompactsMB{}.

The bottom row of Fig.~\ref{fig:FirstProfile} shows the evolution of the stellar density profile. Compared to the main progenitors of central \Normals\ (solid orange), the main progenitors of central \CompactsSB{} (solid green) have lower stellar densities at 1 kpc at all times, but have steeper inner stellar density profiles, catching up with the density of the \Normals\ at the resolution limit of $\approx$300 pc.  \CompactsMB{} (solid blue) have similar  stellar densities to \Normals\ (solid orange) at all radii before $z = 0.5$, when they start to have higher stellar density in the inner region.

Figure \ref{fig:FirstProfile} reveals that  the concentrated star formation of \Compacts{} is the consequence of stronger gas infall, which in turn is due to its lower angular momentum. We see in Sects.~\ref{subsec:merger} and \ref{subsec:environment} that \Compacts{} tend to undergo fewer mergers and interactions and also live in lower mass halos, with fewer satellites. Therefore, the environment  can have an influence on the angular momentum of dwarf galaxies: a denser environment and more galaxy mergers lead to different gas distributions, where lower angular momentum does not accumulate in the inner regions.

\section{What  physical mechanisms make TNG50 galaxies compact?}
\label{sec:details}

We now explore the different  physical mechanisms leading to \Compact\ galaxies in TNG50.

\subsection{Mergers of globular clusters?}

If \Compacts\ were formed by the mergers of globular clusters, we would expect to see an enhanced merger rate at the epochs of globular cluster mergers. However, the median ex situ fraction of stellar mass is lower in \Compacts\ than in \Normals\ at all epochs (dashed curves in the top panel of Fig.~\ref{fig:mergeevol}). Thus, at best, only a small minority of \Compacts\ could be formed by mergers of globular clusters.

\subsection{Initial compactness?}

If central \Compacts\ were initially small upon formation of their most massive progenitors, one would see this in the median trends of main progenitor size. However, the early ($z>4$) sizes of the main progenitors of central \Compacts\ are a good match to those of \Normals\ (Fig.~\ref{fig:sizemassevol}). Therefore, the origin of \Compacts\ in TNG50 cannot be ascribed to initial compactness.

\subsection{Active galactic nuclei?}

\label{sec:bhagn}

In TNG50, a BH is seeded in a subhalo  only if its halo mass exceeds $7.3 \times 10^{10} \msun$  \citep{Weinberger2017MNRAS}. Therefore, central galaxies that have a lower mass-accretion rate will have more difficulty in hosting a BH. During galaxy evolution, this accretion can be produced by mergers and interactions, which help a galaxy to reach the TNG halo mass threshold that needs to be met in order to grow a BH.

\begin{figure}[htpb!]
    \center
    \includegraphics[width=\hsize]{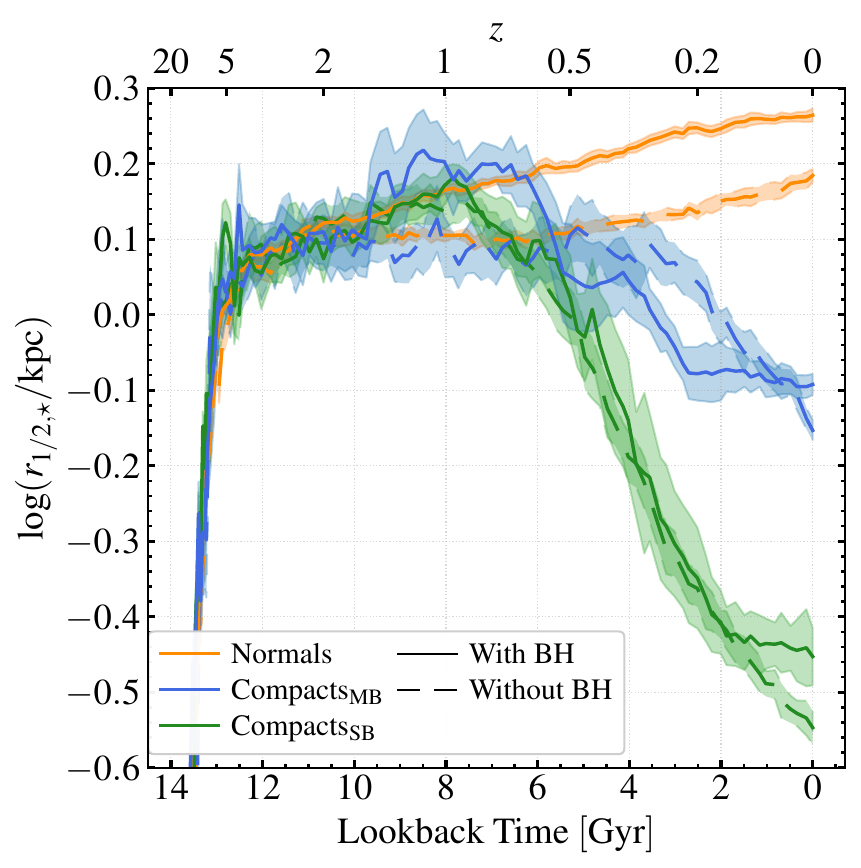}
    \caption{Median evolution of half-mass radius for central galaxies with a BH at $z=0$  ({solid lines}) or that never had a BH ({dashed lines}). The colors correspond to those in Fig.~\ref{fig:sizemassevol}.
   }
  \label{fig:CentralBH}
\end{figure}

At $z=1$, the BH occupation fractions of central \Compacts\ are much smaller than for \Normals: 0\%\ and 8\% of \CompactsSB{} and \CompactsMB{} have BHs in contrast to 31\% of \Normals. By $z$=0, the BH occupation fractions evolve to $22\%$, $49\%,$ and $60\%$ for \CompactsSB, \CompactsMB, and  \Normals, respectively. These lower occupation fractions are due the lower halo masses of dwarf galaxies, as seen in Fig.~\ref{fig:envtevol} (even \Normals{} have a median $z = 0$ halo mass that is relatively close to the threshold for seeding a BH).
 
There are three classes of BH occupation: galaxies that never hosted a BH, galaxies that have a BH at $z=0$, and galaxies that had a BH in the past and lost it. We  discard this latter class of BH occupation, and refer to the first and second ones as ``without BH'' and ``with  BH'', respectively. 

Figure \ref{fig:CentralBH} shows the evolution of size split by $z$=0 BH occupation. The size evolution is only slightly altered for galaxies with BHs. The sizes of \Normals\ are significantly larger if they end up with BHs than if they do not, as expected. However, for \Compacts, the $z$=0 sizes are only weakly affected by the presence of BHs. Therefore, the absence of the BH, and henceforth  the absence of AGN feedback cannot be the main driver of the production of \Compacts{}. Indeed, comparing several nonpublic higher-resolution TNG simulations with different subgrid physics, \citeauthor{Pillepich2018MNRAS} (\citeyear{Pillepich2018MNRAS}, lower-right panel of their fig.~8) found that BH feedback has no influence on the sizes of low-mass galaxies at $z = 0$.

\subsection{Initial halo spin?}

Analyzing the Illustris-5 simulation, \cite{Rodriguez2017MNRAS} suggested that the angular momentum of the halo is the key parameter defining the morphology of a dwarf galaxy. The second rows of Fig.~\ref{fig:FirstProfile} shows that, at $z=1.5$ and $z = 1$, before the stellar compaction, the gas angular momentum profiles of \Compacts\ are not so different from those of  \Normals{}. The difference in gas spin of \Compacts{} becomes apparent at $z = 0.5$, especially for the inner regions. At this time, \Compacts{} already have smaller sizes than \Normals{}, which shows that the initial halo spin is not the deciding factor causing \Compacts{} to become compact, although the decrease in size is related to the gas distribution in these galaxies. 

\subsection{Lack of mergers and rapid interactions?}

In Sect.~\ref{subsec:merger}, we show that mergers contribute only a few percent to the $z$=0 stellar mass of all three size classes of galaxies (3\% for \Normals\ and $\sim$2\% for \Compacts; see the top panel of Fig.~\ref{fig:mergeevol}). Also, \Compacts\ acquire their ex situ mass  earlier, and after $z\sim 2$ their ex situ mass growth is slower than that of \Normals\ as they evolve ``passively'' through accretion and in situ star formation. These results  suggest that the lack of mergers plays a role in the formation of the \Compact{} galaxies.

\begin{figure}[htpb!]
    \centering
\includegraphics[width=1\hsize]{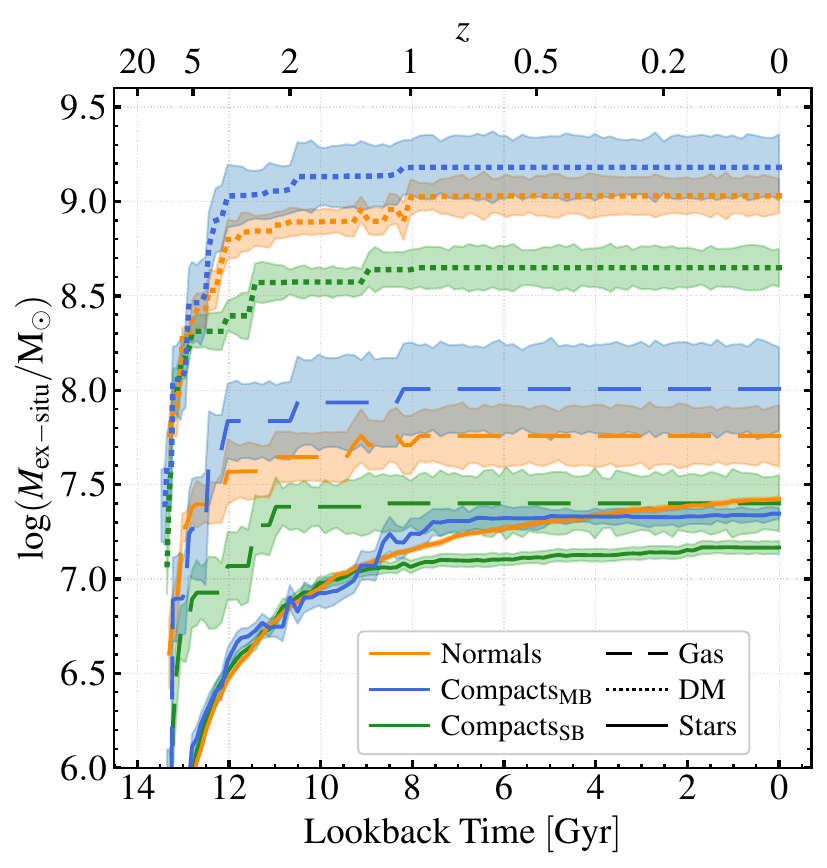}
    \caption{Same as the bottom panel of Fig.~\ref{fig:mergeevol}, but for
      the total ex situ mass, split into the three components. Only the 30 galaxies considered in Fig.~\ref{fig:FirstProfile} are shown here.
     }
  \label{fig:ExSituTot}
\end{figure}

For the same galaxies as Fig.~\ref{fig:FirstProfile}, we computed the  ex situ contributions for the gas and DM components, which are not provided in the TNG supplementary catalogs. Figure \ref{fig:ExSituTot} shows  that starting at $z=5$, \CompactsSB{} have only half the ex situ gas and DM mass of the \CompactsMB\ and \Normals, and this ratio remains until $z=0$. On the other hand,  the ex situ masses of all three components of \CompactsMB{} are similar to those of the \Normals. This  again reveals a difference between the two size classes of \Compacts{}. However, while \CompactsMB{} have a similar ex situ mass to \Normals, mergers stopped contributing earlier to their mass growth, as we see in Fig.~\ref{fig:histMergers}.

\subsection{Low-angular-momentum mergers?}

Mergers may prevent galaxies from growing in size if the specific orbital angular momentum of the merging galaxies is greater than the internal specific angular momentum of the two galaxies. We checked whether or not the progenitors of the present-day \Compacts\ suffered early from lower-orbital-angular-momentum encounters than \Normals, analyzing the same random galaxies used in Fig.~\ref{fig:FirstProfile}. 

Then, through their merger trees, we computed the angular momentum of the system that encloses the main progenitor and the secondary galaxy. To this end, we computed the importance of the orbital angular momentum of the merging system using 
\begin{equation}
    \frac{j_\mathrm{merger}}{j_\mathrm{max}} = \frac{\left|\vec{r_{12}} \times \vec{v_{12}} \right|}{\left|\vec{r_{12}}\right|\left|\vec{v_{12}}\right|},
\end{equation}
where  $j_\mathrm{merger}$ is the specific angular momentum of the system, $j_\mathrm{max}$ is the maximum value, while $\vec{r_{12}}$ and $\vec{v_{12}}$ are the position and velocity in the main progenitor frame, respectively. Values close to unity indicate very high-angular-momentum mergers. We do not find any statistically difference between the median values for $j_\mathrm{merger}/j_\mathrm{max}$, finding  $0.86$ and $0.87$  for \CompactsSB{} and \CompactsMB{}, respectively, while \Normals{} have a median of $0.88$. Thus,  mergers tend to be high angular momentum regardless of the galaxy size class.  This indicates that rather than the merger properties, it is the merger frequency and contribution that are the key drivers of the growth of \Compacts{}, as discussed in Sect.~\ref{subsec:merger}.

\subsection{Low-angular-momentum-gas infall?}
\label{subsec:lowerang}

The top panels of Fig.~\ref{fig:FirstProfile} clearly indicate that \Compacts{} tend to have inner gas infall (which is particularly intense for \CompactsSB{}), leading to concentrated star formation, and then to a young stellar core, making the galaxy compact. 

\begin{figure}[htpb!]
    \centering
\includegraphics[width=1 \hsize]{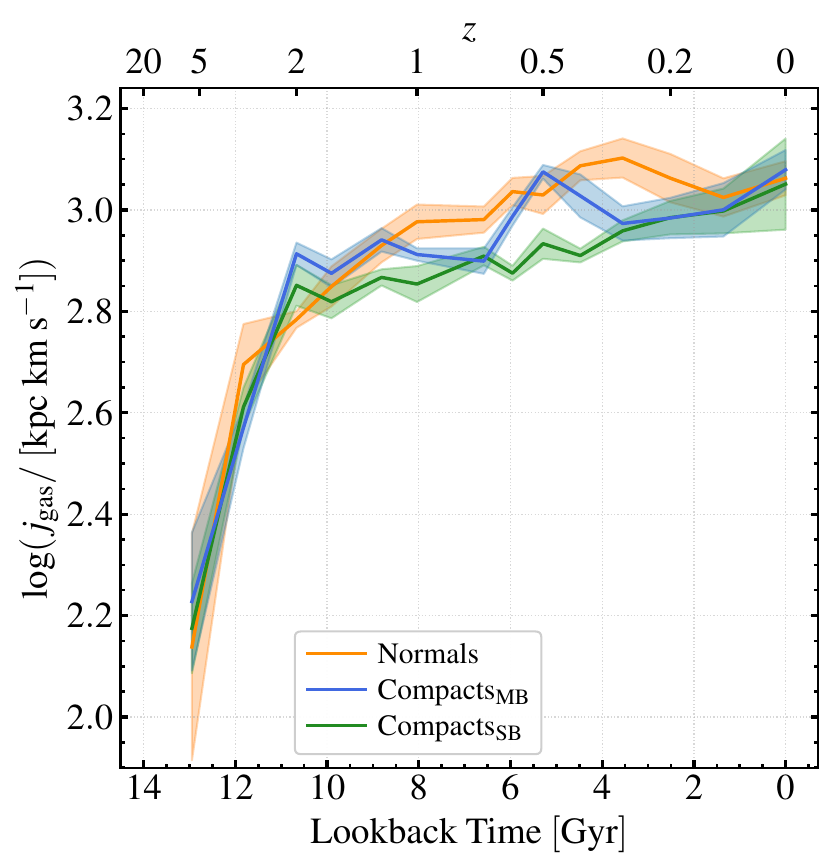}
    \caption{Same as Fig.~\ref{fig:sizemassevol}, but for the angular
      momentum of the current accreting gas (negative radial velocity, $v_r < 0$ km s$^{-1}$) measured in the region above two times the half-gas-mass radius ($r > 2 r_{1/2, \mathrm{gas}}$). Only the 30 galaxies considered in Fig.~\ref{fig:FirstProfile} are shown here, for 13 epochs.
 }
  \label{fig:AngMomGas}
\end{figure}

More precisely, gas infall leads to the accumulation of  lower-angular-momentum gas  in the center (second row of panels of Fig.~\ref{fig:FirstProfile}). This  suggests that  gas infall is caused by the accretion
 of lower-angular-momentum gas onto the galaxy. To verify this, Fig.~\ref{fig:AngMomGas} shows the angular momentum evolution for the outer (recent infalling) gas particles: $v_{r} < 0$ km s$^{-1}$ and $r > 2 r_{1/2, \mathrm{gas}}$.

We computed the median angular momentum for the same 30 central galaxies considered in Fig.~\ref{fig:FirstProfile} at 13 snapshots, roughly regularly spaced in time. Figure~\ref{fig:AngMomGas} indicates that, from $z=1$ (5.5  Gyr lookback time) to $z=0.25$ (3 Gyr lookback time), the gas  infalling onto \Compacts\ has a lower angular momentum than that infalling onto \Normals, except for central \CompactsMB\ at one epoch ($z=0.5$). This time interval corresponds to the epoch of decreasing size for \Compacts{}. Therefore, the infall of low-angular-momentum gas appears to be linked to the compaction of dwarf galaxies.

\section{Discussion}
\label{Discussion}

\subsection{Comparison to previous work}

First, our findings agree with those of \cite{Deeley+23} that \Compacts\ ending up as centrals preferentially form stars in their inner regions. However, while these authors explain that these \Compacts\ grew by continuous growth, they do not explain why they became compact.

We now compare our main conclusion  with other previous works, namely that \Compact\ dwarf galaxies that are centrals (or isolated) shrunk as a result of a lack of mergers between  $z\sim2$  and  $z\sim0$ , allowing gas infall, inner star formation, and the buildup of a stellar core.

Several studies point to different scenarios for massive galaxies. In a series of articles, A. Dekel and collaborators \citep{Dekel2014MNRAS,Zolotov2015MNRAS,Lapiner2023MNRAS} argued that massive galaxies at $z\sim2$ went through the following steps: their gas disks, fed by powerful accreting streams and minor mergers, became unstable, creating important clumps that then spiraled into the center by dynamical friction, where they merged with the central one, and  building up a compact star forming galaxy that they coined a ``blue nugget''; these blue nuggets passively evolve into ``red nuggets''. Mergers of gas-rich dwarf galaxies were proposed as the mechanism forming blue compact dwarfs \citep[e.g.,][]{Bekki2008MNRAS, Watts2016MNRAS}. The merger will lead to a central gas concentration that results in starburst.

However, mergers are rare for dwarf galaxies  \citep{Cattaneo+11}. The lack of mergers also suppresses the minor mergers as one avenue to generate the violent disk instability mechanism outlined above. The TNG50 simulations point to the opposite scenario:  those \Compacts\ that end up as centrals  are caused by a {lack} of mergers, allowing gas infall to produce efficient star formation in the inner regions.

Although central \Compacts{} have gas infall (Fig.~\ref{fig:FirstProfile}), they do not show gas contraction, contrary to  the red nuggets scenario. This  suggests that  gas infall maintains the reservoir for inner star formation without concentrating gas in the galaxy cores.

Finally, \cite{Lohmann+23} studied the formation of massive compact galaxies (MCGs) in TNG100 and showed that these galaxies formed by the accretion of low-angular momentum gas. Although this scenario is similar to central \Compact\ dwarf population evolution (Sect.~\ref{subsec:lowerang}), the dwarf galaxies shrink from $z=1$ onward in contrast to the MCGs, which do not shrink, but experience much slower growth in size than other high-mass galaxies. 

\subsection{Are our results caused by the limitations of the simulation?}

\subsubsection{Resolution effects}
\label{sec:resolution}
It is possible that the 288~pc softening scale of the star and DM particles at $z<1$ (and constant comoving resolution at earlier epochs) could severely affect our results. As the gas resolution is much finer, if stars form in a thin disk, they will later diffuse  into a thicker disk. This diffusion occurs  in a numerical two-body relaxation time, which may be faster or slower than that observed or expected from different diffusion mechanisms (molecular clouds, \citealt{Spitzer&Schwarzschild51}; spiral arms, \citealt{Barbanis&Woltjer67}; black holes, \citealt{Lacey&Ostriker85}; minor mergers, \citealt{Toth&Ostriker92}; ingested satellites, \citealt*{Quinn+93}; and DM subhalos, \citealt{Font+01}).

\begin{figure}[ht]
  \centering
  \includegraphics[width=\hsize]{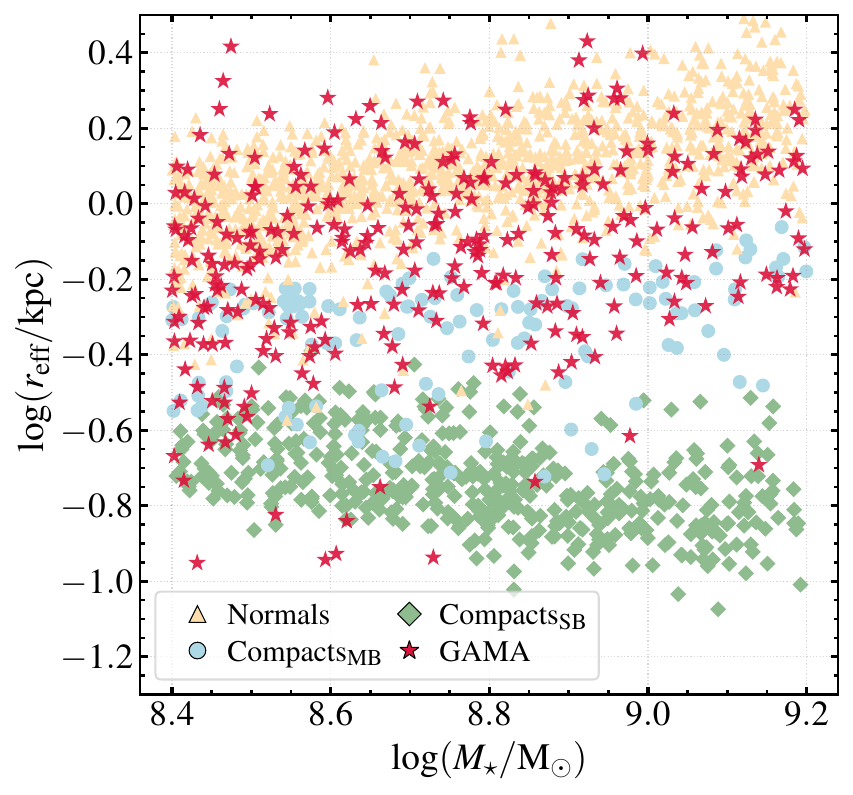}
  \caption{
Present-day half-projected-light radius vs. stellar mass relation, for GAMA and TNG50 galaxies.  The {triangles}, {circles,} and {diamonds}  
      distinguish the different samples, respectively: \Normals, \CompactsMB,\ and \CompactsSB, while the
      {stars} represent all the galaxies in GAMA. 
}
   \label{fig:MassSIzeGAMA}
\end{figure}

\cite{Pillepich2019MNRAS} argued that the sizes of galaxies with $M_\star > 10^8 \msun$ are correctly resolved in TNG50. However, its limited resolution may explain why our \Compacts\ are much bigger than compact ellipticals and ultracompact dwarfs in our range of stellar masses ($8.4 < \lmsun < 9.2$).

We compared the TNG50 dwarfs to similar-mass dwarfs from a complete observational sample: the fourth data release of the Galaxy and Mass Assembly (GAMA) survey \citep{Driver2022MNRAS}. We computed the sizes by taking the galaxy redshifts and masses from the GAMA \texttt{StellarMassesv19} table, and the $i$-band effective (projected half-light) radii from the single-S\'ersic fits provided in the \texttt{SersicCatSDSSv09} table.  We used the \texttt{MagPhysv06} table for the specific star formation rates.  We selected the galaxies between $z = 0.002$ and $z = 0.025$ in our stellar mass range ($8.2 < \lmsun < 9.4$), restricting our sample to galaxies whose log size error is smaller than 0.1. To compare with the TNG50 galaxies, we used the TNG50 supplementary data catalog {\sf Stellar Projected Sizes}  \citep{Genel2018MNRAS} for the $z$=0 $i$-band effective radii at a random orientation.

Figure \ref{fig:MassSIzeGAMA} shows the relation between effective radius and stellar mass for both the TNG50 galaxies and those observed in the GAMA survey.  TNG50 reproduces the observed  SMR for \Normals{} and \CompactsMB{} galaxies fairly well.  But we note that for galaxies with stellar mass with $8.4 < \lmsun < 8.8$, the GAMA survey shows a greater number of very compact galaxies than the TNG50, where the smallest effective radius is $\sim 100$~pc. Indeed, the \CompactsSB\ in TNG50 are at the size resolution of TNG50 at $z=0$ (black lines in the upper panels of Fig.~\ref{fig:indiv_compact}) and would  probably have been smaller in a similar simulation with better spatial resolution.

\subsubsection{The secondary branch}

The existence of the clearly distinguished secondary branch may be caused by a physical process causing runaway stellar compaction. Identifying this process is not straightforward.

We note several differences between \CompactsMB\ and \CompactsSB:
  (1) the former tend to end up as satellites, while the latter tend to end up as centrals (Table~\ref{tab:samples});
  (2) the \CompactsSB{} have  lower ex situ fractions than the \CompactsMB\ (Fig.~\ref{fig:ExSituTot});
  and (3) the preference for inner star formation is much more pronounced in the \CompactsSB{} (bottom panel of Fig.~\ref{fig:ssfrevol}).

Therefore, we conclude that \CompactsSB{} are extreme cases of  \Compacts{} in terms of  concentrated star formation. This concentrated star formation leads to an anticorrelation between size and stellar mass: galaxies with higher stellar mass (that form more stars in their inner regions) have smaller sizes. This is what we see clearly in the secondary branch (Fig.~\ref{fig:sizevsmass_classes}).

The intensity of the inner star formation depends on the inner gas infall (top panels of Fig.~\ref{fig:FirstProfile}), which is controlled by the importance of major and intermediate mergers: below a certain level of importance there is uncontrolled inner gas infall.
  
  The small sizes of the stellar components of \CompactsSB{} are at the limit of the spatial resolution (softening scale) of TNG50. With better resolution, the secondary branch of the SMR would  extend to lower sizes and thus not be so sharp.  However, we are not presently able to explain the gap between the two branches, beyond our finding that the  \CompactsSB{} are the result of intense concentrated star formation.
\begin{figure}[ht]
  \centering
  \includegraphics[width=\hsize]{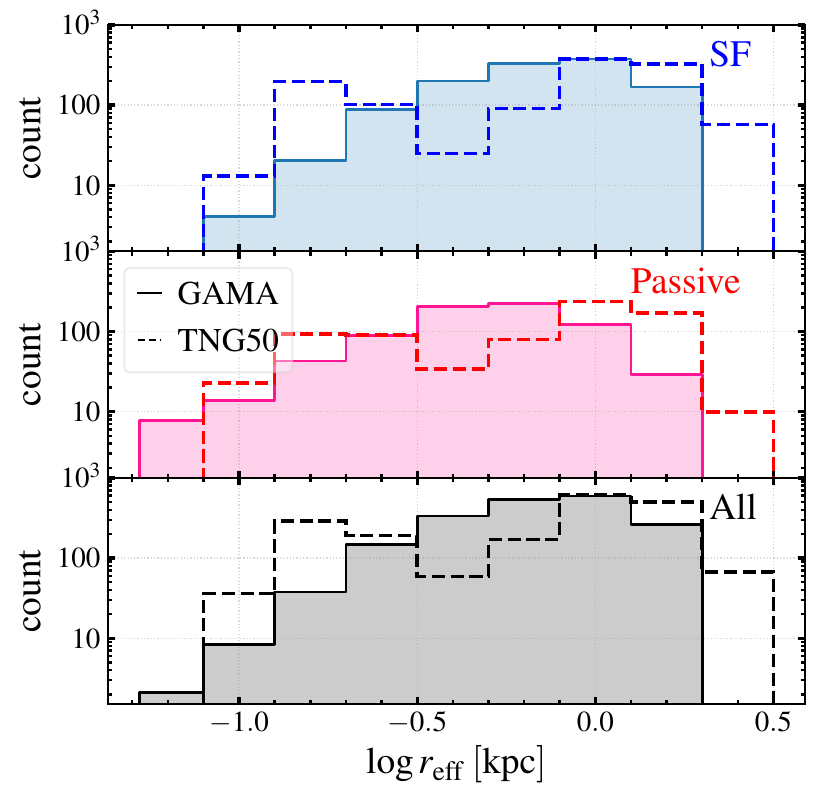}
  \caption{Normalized size distributions for GAMA and TNG50 galaxies, split between star forming ({blue}, \textbf{top}) and passive ({red}, \textbf{center}). The \textbf{bottom} panel shows the distribution for all galaxies.  Each GAMA galaxy was weighted in inverse proportion to the volume up to which it can be seen. 
}
   \label{fig:GAMA}
   \end{figure}

Figure \ref{fig:MassSIzeGAMA} shows a possible gap in the GAMA SMR at effective radii that increase from 140~pc to 450~pc ($\log(r_\mathrm{eff}/\mathrm{kpc})$ between $\simeq -0.85$ and $-0.35$) between $\log (M_\star/{\rm M}_\odot)=8.4$ and 9.2, but it is not as pronounced  as in TNG50. 
This gap in GAMA is also seen as the abrupt rise  in Fig.~\ref{fig:GAMA} of the GAMA counts at $\log(r_{\rm eff}/{\rm kpc}) \simeq -0.7$ (200 pc). The histogram of log sizes weakens the observed gap, the location of which in the SMR increases slightly with stellar mass, as noted above. However, a Hartigan dip test \citep{Hartigan85} of the residuals of $\log r_{1/2}$ indicates that this gap is not statistically significant. Furthermore, it is less pronounced if we increase the maximum redshift for our sample of GAMA dwarfs.

\subsubsection{High star formation rates in the cores of \Compacts}

It is difficult to understand why the simulated \Compacts\ tend to be more highly star forming than \Normals, while the opposite trend appears in the SDSS \citep{Shen2003MNRAS}. Figure~\ref{fig:GAMA} shows that the star forming galaxies  show a  bimodality in TNG50 that is not present in GAMA. Also, passive galaxies are likely to have smaller sizes in GAMA than in TNG50.

One possible explanation for this is that TNG50 is missing low-mass BHs in dwarf galaxies in general and in the \CompactsSB{} in particular. More precisely, the minimum halo mass to seed a BH is $7.3 \times10^{10}\msun$ in the TNG simulations, which may be too large (at least in TNG50) and lead to excessively low BH occupation fractions, especially for \Compacts{} that have a lower contribution from mergers and therefore have more difficulty reaching the threshold to seed a BH.  

As seen in Fig.~\ref{fig:ssfrevol}, not only are \Compacts\  bluer (they have higher sSFR) than \Normals, but they have blue cores (their sSFR is higher in their cores). Although the presence or absence of the BH does not affect the decrease in the size of \Compacts{} (Sect.~\ref{sec:bhagn}), central feedback from the BH could perhaps regulate this higher sSFR. Both TNG50 star formation and BH prescriptions may favor concentrated star formation and thus the formation of compact dwarf galaxies.

\section{Conclusions}
\label{Conclusion}

In this first study analyzing the TNG50 cosmological hydrodynamical  simulation to explore which mechanisms regulate the corpulence of galaxies, we explored the physical mechanisms that cause  central dwarf galaxies  to become \Compact{}. We selected dwarf galaxies according to their $z$=0 properties: stellar log masses were between 8.4 and 9.2 (solar units). We called to ``Compacts'' those galaxies lying in the lower envelope of the main branch of the size--mass relation (``\CompactsMB{}'') ---easily selected using Eqs. (\ref{eq:01}) and (\ref{eq:02})--- as well as the dwarfs in the secondary branch,
with a stellar half-mass radius of below 450 pc (``\CompactsSB''), while the galaxies that ended with larger sizes are referred to as ``\Normals''. We furthermore restricted our samples to the \Compacts\ and \Normals\ that ended up as centrals of their hosts.

Galaxies that end up as central \Compacts{} typically have a similar size evolution to \Normals\ until $z=0.8$, at which point they start shrinking, while the \Normals{} continue to grow (Figs.~\ref{fig:sizemassevol}).  Central \Compacts\ live in lower density environments (Figs.~\ref{fig:envtevol}), and  have lower  ex situ stellar mass fractions (Figs.~\ref{fig:mergeevol}, \ref{fig:ExSituTot}, \ref{fig:CentralBH}). These are the galaxies that suffered fewer major and intermediate mergers before $z=1$ (Figs.~\ref{fig:mergeevol} and \ref{fig:histMergers})  and accreted more lower angular momentum gas (Fig.~\ref{fig:AngMomGas}).

These mergers would otherwise pump their orbital angular momentum into the remnant and prevent radial gas inflow. Here, the gas infall is ubiquitous and produces  concentrated star formation (Figs.~\ref{fig:ssfrevol} and \ref{fig:FirstProfile}), which leads to the formation of a stellar core and makes the galaxy more spherical (Fig. \ref{fig:shapeevol}).  While this lack of mergers prevents the formation and growth of BHs ---whose AGN could prevent the gas infall---, the rare presence or not of a BH has little effect on the evolution of the stellar sizes of \Compacts, in contrast to the situation for \Normals\  (Fig.~\ref{fig:CentralBH}). The compaction of individual central \Compacts\ is gradual and occurs when the ratio of inner to outer sSFR is enhanced (Fig.~\ref{fig:indiv_compact}). Their gas infall (much stronger in \CompactsSB{} than in \CompactsMB{}) occurs while their gas half-mass-radii keep growing (Fig.~\ref{fig:sizemassevol}).

It is not clear how our conclusions depend on the limitations of the TNG50 simulation. Observed \Compacts\  from the GAMA survey also show weak but insignificant signs of bimodality in size (Fig.~\ref{fig:MassSIzeGAMA}). However, while the \CompactsSB{} are typically star forming ---with ``blue'' cores (Fig.~\ref{fig:ssfrevol}), as observed in some compact dwarfs--- and appear round (Fig.~\ref{fig:shapeevol}), GAMA \Compacts\ tend to be passive (Fig.~\ref{fig:GAMA}). This difference may be caused by the absence of BHs in very low-mass halos in the subgrid physics of TNG50. Therefore, a similar  study to the present one ought to be repeated with current (e.g., NewHorizon, \citealt{Dubois+21} and FIRE-2, \citealt{Wetzel+23}) and future high-resolution cosmological hydrodynamical simulations.

\section*{Acknowledgements}
We thank the anonymous referee for useful comments. APA thanks the São Paulo Research Foundation, FAPESP, for
financial support through contracts 2022/05059-2 and  2020/16152-8. GBLN
acknowledges partial financial support from CNPq grant 303130/2019-9.

\bibliography{ref}

\begin{thebibliography}{61}
\expandafter\ifx\csname natexlab\endcsname\relax\def\natexlab#1{#1}\fi

\bibitem[{{Barbanis} \& {Woltjer}(1967)}]{Barbanis&Woltjer67}
{Barbanis}, B. \& {Woltjer}, L. 1967, \apj, 150, 461

\bibitem[{{Barnes} \& {Hernquist}(1992)}]{Barnes1992ARAA}
{Barnes}, J.~E. \& {Hernquist}, L. 1992, \araa, 30, 705

\bibitem[{{Barnes} \& {Hernquist}(1991)}]{Barnes1991ApJ}
{Barnes}, J.~E. \& {Hernquist}, L.~E. 1991, \apjl, 370, L65

\bibitem[{{Bekki}(2008)}]{Bekki2008MNRAS}
{Bekki}, K. 2008, \mnras, 388, L10

\bibitem[{{Bernardi} {et~al.}(2011){Bernardi}, {Roche}, {Shankar}, \& {Sheth}}]{Bernardi+11}
{Bernardi}, M., {Roche}, N., {Shankar}, F., \& {Sheth}, R.~K. 2011, \mnras, 412, 684

\bibitem[{{Brodie} {et~al.}(2011){Brodie}, {Romanowsky}, {Strader}, \& {Forbes}}]{Brodie2011AJ}
{Brodie}, J.~P., {Romanowsky}, A.~J., {Strader}, J., \& {Forbes}, D.~A. 2011, \aj, 142, 199

\bibitem[{{Cattaneo} {et~al.}(2011){Cattaneo}, {Mamon}, {Warnick}, \& {Knebe}}]{Cattaneo+11}
{Cattaneo}, A., {Mamon}, G.~A., {Warnick}, K., \& {Knebe}, A. 2011, \aap, 533, A5

\bibitem[{{Chilingarian} \& {Mamon}(2008)}]{Chilingarian&Mamon08}
{Chilingarian}, I.~V. \& {Mamon}, G.~A. 2008, \mnras, 385, L83

\bibitem[{{Dashyan} {et~al.}(2019){Dashyan}, {Choi}, {Somerville}, {Naab}, {Quirk}, {Hirschmann}, \& {Ostriker}}]{Dashyan+19}
{Dashyan}, G., {Choi}, E., {Somerville}, R.~S., {et~al.} 2019, \mnras, 487, 5889

\bibitem[{{Davis} {et~al.}(1985){Davis}, {Efstathiou}, {Frenk}, \& {White}}]{Davis1985ApJ}
{Davis}, M., {Efstathiou}, G., {Frenk}, C.~S., \& {White}, S.~D.~M. 1985, \apj, 292, 371

\bibitem[{{Deeley} {et~al.}(2023){Deeley}, {Drinkwater}, {Sweet}, {Bekki}, {Couch}, \& {Forbes}}]{Deeley+23}
{Deeley}, S., {Drinkwater}, M.~J., {Sweet}, S.~M., {et~al.} 2023, \mnras, 525, 1192

\bibitem[{{Dekel} \& {Burkert}(2014)}]{Dekel2014MNRAS}
{Dekel}, A. \& {Burkert}, A. 2014, \mnras, 438, 1870

\bibitem[{{Dekel} {et~al.}(2009){Dekel}, {Sari}, \& {Ceverino}}]{Dekel+09}
{Dekel}, A., {Sari}, R., \& {Ceverino}, D. 2009, \apj, 703, 785

\bibitem[{{Dekel} \& {Silk}(1986)}]{Dekel1986ApJ}
{Dekel}, A. \& {Silk}, J. 1986, \apj, 303, 39

\bibitem[{{Driver} {et~al.}(2022){Driver}, {Bellstedt}, {Robotham}, {Baldry}, {Davies}, {Liske}, {Obreschkow}, {Taylor}, {Wright}, {Alpaslan}, {Bamford}, {Bauer}, {Bland-Hawthorn}, {Bilicki}, {Bravo}, {Brough}, {Casura}, {Cluver}, {Colless}, {Conselice}, {Croom}, {de Jong}, {D'Eugenio}, {De Propris}, {Dogruel}, {Drinkwater}, {Dvornik}, {Farrow}, {Frenk}, {Giblin}, {Graham}, {Grootes}, {Gunawardhana}, {Hashemizadeh}, {H{\"a}u{\ss}ler}, {Heymans}, {Hildebrandt}, {Holwerda}, {Hopkins}, {Jarrett}, {Heath Jones}, {Kelvin}, {Koushan}, {Kuijken}, {Lara-L{\'o}pez}, {Lange}, {L{\'o}pez-S{\'a}nchez}, {Loveday}, {Mahajan}, {Meyer}, {Moffett}, {Napolitano}, {Norberg}, {Owers}, {Radovich}, {Raouf}, {Peacock}, {Phillipps}, {Pimbblet}, {Popescu}, {Said}, {Sansom}, {Seibert}, {Sutherland}, {Thorne}, {Tuffs}, {Turner}, {van der Wel}, {van Kampen}, \& {Wilkins}}]{Driver2022MNRAS}
{Driver}, S.~P., {Bellstedt}, S., {Robotham}, A. S.~G., {et~al.} 2022, \mnras, 513, 439

\bibitem[{{Dubois} {et~al.}(2021){Dubois}, {Beckmann}, {Bournaud}, {Choi}, {Devriendt}, {Jackson}, {Kaviraj}, {Kimm}, {Kraljic}, {Laigle}, {Martin}, {Park}, {Peirani}, {Pichon}, {Volonteri}, \& {Yi}}]{Dubois+21}
{Dubois}, Y., {Beckmann}, R., {Bournaud}, F., {et~al.} 2021, \aap, 651, A109

\bibitem[{{Font} {et~al.}(2001){Font}, {Navarro}, {Stadel}, \& {Quinn}}]{Font+01}
{Font}, A.~S., {Navarro}, J.~F., {Stadel}, J., \& {Quinn}, T. 2001, \apjl, 563, L1

\bibitem[{{Furlong} {et~al.}(2017){Furlong}, {Bower}, {Crain}, {Schaye}, {Theuns}, {Trayford}, {Qu}, {Schaller}, {Berthet}, \& {Helly}}]{Furlong2017MNRAS}
{Furlong}, M., {Bower}, R.~G., {Crain}, R.~A., {et~al.} 2017, \mnras, 465, 722

\bibitem[{{Genel} {et~al.}(2015){Genel}, {Fall}, {Hernquist}, {Vogelsberger}, {Snyder}, {Rodriguez-Gomez}, {Sijacki}, \& {Springel}}]{Genel2015ApJ}
{Genel}, S., {Fall}, S.~M., {Hernquist}, L., {et~al.} 2015, \apjl, 804, L40

\bibitem[{{Genel} {et~al.}(2018){Genel}, {Nelson}, {Pillepich}, {Springel}, {Pakmor}, {Weinberger}, {Hernquist}, {Naiman}, {Vogelsberger}, {Marinacci}, \& {Torrey}}]{Genel2018MNRAS}
{Genel}, S., {Nelson}, D., {Pillepich}, A., {et~al.} 2018, \mnras, 474, 3976

\bibitem[{{Gunn} \& {Gott}(1972)}]{Gunn1972ApJ}
{Gunn}, J.~E. \& {Gott}, J.~Richard, {\rm III}. 1972, \apj, 176, 1

\bibitem[{{Guo} \& {White}(2008)}]{Guo&White08}
{Guo}, Q. \& {White}, S.~D.~M. 2008, \mnras, 384, 2

\bibitem[{{Hartigan} \& {Hartigan}(1985)}]{Hartigan85}
{Hartigan}, J.~A. \& {Hartigan}, P.~M. 1985, The Annals of Statistics, 13, 70

\bibitem[{{Haslbauer} {et~al.}(2019){Haslbauer}, {Dabringhausen}, {Kroupa}, {Javanmardi}, \& {Banik}}]{Haslbauer+19}
{Haslbauer}, M., {Dabringhausen}, J., {Kroupa}, P., {Javanmardi}, B., \& {Banik}, I. 2019, \aap, 626, A47

\bibitem[{{Kroupa}(1998)}]{Kroupa1998MNRAS}
{Kroupa}, P. 1998, \mnras, 300, 200

\bibitem[{{Lacey} \& {Ostriker}(1985)}]{Lacey&Ostriker85}
{Lacey}, C.~G. \& {Ostriker}, J.~P. 1985, \apj, 299, 633

\bibitem[{{Lapiner} {et~al.}(2023){Lapiner}, {Dekel}, {Freundlich}, {Ginzburg}, {Jiang}, {Kretschmer}, {Tacchella}, {Ceverino}, \& {Primack}}]{Lapiner2023MNRAS}
{Lapiner}, S., {Dekel}, A., {Freundlich}, J., {et~al.} 2023, \mnras, 522, 4515

\bibitem[{{Lohmann} {et~al.}(2023){Lohmann}, {Schnorr-M{\"u}ller}, {Trevisan}, {Ricci}, \& {Clerici}}]{Lohmann+23}
{Lohmann}, F.~S., {Schnorr-M{\"u}ller}, A., {Trevisan}, M., {Ricci}, T.~V., \& {Clerici}, K.~S. 2023, \mnras, 524, 5266

\bibitem[{{Mahani} {et~al.}(2021){Mahani}, {Zonoozi}, {Haghi}, {Je{\v{r}}{\'a}bkov{\'a}}, {Kroupa}, \& {Mieske}}]{Mahani2021MNRAS}
{Mahani}, H., {Zonoozi}, A.~H., {Haghi}, H., {et~al.} 2021, \mnras, 502, 5185

\bibitem[{{Mamon}(1987)}]{Mamon87}
{Mamon}, G.~A. 1987, \apj, 321, 622

\bibitem[{{Marinacci} {et~al.}(2018){Marinacci}, {Vogelsberger}, {Pakmor}, {Torrey}, {SpringMarinacciel}, {Hernquist}, {Nelson}, {Weinberger}, {Pillepich}, {Naiman}, \& {Genel}}]{Marinacci2018MNRAS}
{Marinacci}, F., {Vogelsberger}, M., {Pakmor}, R., {et~al.} 2018, \mnras, 480, 5113

\bibitem[{{Merritt}(1983)}]{Merritt1983ApJ}
{Merritt}, D. 1983, \apj, 264, 24

\bibitem[{{Mieske} {et~al.}(2002){Mieske}, {Hilker}, \& {Infante}}]{Mieske2002AA}
{Mieske}, S., {Hilker}, M., \& {Infante}, L. 2002, \aap, 383, 823

\bibitem[{{Mieske} {et~al.}(2012){Mieske}, {Hilker}, \& {Misgeld}}]{Mieske2012AA}
{Mieske}, S., {Hilker}, M., \& {Misgeld}, I. 2012, \aap, 537, A3

\bibitem[{{Misgeld} \& {Hilker}(2011)}]{Misgeld2011MNRAS}
{Misgeld}, I. \& {Hilker}, M. 2011, \mnras, 414, 3699

\bibitem[{{Moore} {et~al.}(1996){Moore}, {Katz}, {Lake}, {Dressler}, \& {Oemler}}]{Moore1996Natur}
{Moore}, B., {Katz}, N., {Lake}, G., {Dressler}, A., \& {Oemler}, A. 1996, \nat, 379, 613

\bibitem[{{Naiman} {et~al.}(2018){Naiman}, {Pillepich}, {Springel}, {Ramirez-Ruiz}, {Torrey}, {Vogelsberger}, {Pakmor}, {Nelson}, {Marinacci}, {Hernquist}, {Weinberger}, \& {Genel}}]{Naiman2018MNRAS}
{Naiman}, J.~P., {Pillepich}, A., {Springel}, V., {et~al.} 2018, \mnras, 477, 1206

\bibitem[{{Nelson} {et~al.}(2019){Nelson}, {Pillepich}, {Springel}, {Pakmor}, {Weinberger}, {Genel}, {Torrey}, {Vogelsberger}, {Marinacci}, \& {Hernquist}}]{Nelson2019MNRAS}
{Nelson}, D., {Pillepich}, A., {Springel}, V., {et~al.} 2019, \mnras, 490, 3234

\bibitem[{{Nelson} {et~al.}(2018){Nelson}, {Pillepich}, {Springel}, {Weinberger}, {Hernquist}, {Pakmor}, {Genel}, {Torrey}, {Vogelsberger}, {Kauffmann}, {Marinacci}, \& {Naiman}}]{Nelson2018MNRAS}
{Nelson}, D., {Pillepich}, A., {Springel}, V., {et~al.} 2018, \mnras, 475, 624

\bibitem[{{Pillepich} {et~al.}(2018{\natexlab{a}}){Pillepich}, {Nelson}, {Hernquist}, {Springel}, {Pakmor}, {Torrey}, {Weinberger}, {Genel}, {Naiman}, {Marinacci}, \& {Vogelsberger}}]{Pillepich2018bMNRAS}
{Pillepich}, A., {Nelson}, D., {Hernquist}, L., {et~al.} 2018{\natexlab{a}}, \mnras, 475, 648

\bibitem[{{Pillepich} {et~al.}(2019){Pillepich}, {Nelson}, {Springel}, {Pakmor}, {Torrey}, {Weinberger}, {Vogelsberger}, {Marinacci}, {Genel}, {van der Wel}, \& {Hernquist}}]{Pillepich2019MNRAS}
{Pillepich}, A., {Nelson}, D., {Springel}, V., {et~al.} 2019, \mnras, 490, 3196

\bibitem[{{Pillepich} {et~al.}(2018{\natexlab{b}}){Pillepich}, {Springel}, {Nelson}, {Genel}, {Naiman}, {Pakmor}, {Hernquist}, {Torrey}, {Vogelsberger}, {Weinberger}, \& {Marinacci}}]{Pillepich2018MNRAS}
{Pillepich}, A., {Springel}, V., {Nelson}, D., {et~al.} 2018{\natexlab{b}}, \mnras, 473, 4077

\bibitem[{{Planck Collaboration} {et~al.}(2016){Planck Collaboration}, {Ade}, {Aghanim}, {Arnaud}, {Ashdown}, {Aumont}, {Baccigalupi}, {Banday}, {Barreiro}, {Bartlett}, {Bartolo}, {Battaner}, {Battye}, {Benabed}, {Beno{\^\i}t}, {Benoit-L{\'e}vy}, {Bernard}, {Bersanelli}, {Bielewicz}, {Bock}, {Bonaldi}, {Bonavera}, {Bond}, {Borrill}, {Bouchet}, {Boulanger}, {Bucher}, {Burigana}, {Butler}, {Calabrese}, {Cardoso}, {Catalano}, {Challinor}, {Chamballu}, {Chary}, {Chiang}, {Chluba}, {Christensen}, {Church}, {Clements}, {Colombi}, {Colombo}, {Combet}, {Coulais}, {Crill}, {Curto}, {Cuttaia}, {Danese}, {Davies}, {Davis}, {de Bernardis}, {de Rosa}, {de Zotti}, {Delabrouille}, {D{\'e}sert}, {Di Valentino}, {Dickinson}, {Diego}, {Dolag}, {Dole}, {Donzelli}, {Dor{\'e}}, {Douspis}, {Ducout}, {Dunkley}, {Dupac}, {Efstathiou}, {Elsner}, {En{\ss}lin}, {Eriksen}, {Farhang}, {Fergusson}, {Finelli}, {Forni}, {Frailis}, {Fraisse}, {Franceschi}, {Frejsel}, {Galeotta}, {Galli}, {Ganga}, {Gauthier}, {Gerbino}, {Ghosh}, {Giard},
  {Giraud-H{\'e}raud}, {Giusarma}, {Gjerl{\o}w}, {Gonz{\'a}lez-Nuevo}, {G{\'o}rski}, {Gratton}, {Gregorio}, {Gruppuso}, {Gudmundsson}, {Hamann}, {Hansen}, {Hanson}, {Harrison}, {Helou}, {Henrot-Versill{\'e}}, {Hern{\'a}ndez-Monteagudo}, {Herranz}, {Hildebrandt}, {Hivon}, {Hobson}, {Holmes}, {Hornstrup}, {Hovest}, {Huang}, {Huffenberger}, {Hurier}, {Jaffe}, {Jaffe}, {Jones}, {Juvela}, {Keih{\"a}nen}, {Keskitalo}, {Kisner}, {Kneissl}, {Knoche}, {Knox}, {Kunz}, {Kurki-Suonio}, {Lagache}, {L{\"a}hteenm{\"a}ki}, {Lamarre}, {Lasenby}, {Lattanzi}, {Lawrence}, {Leahy}, {Leonardi}, {Lesgourgues}, {Levrier}, {Lewis}, {Liguori}, {Lilje}, {Linden-V{\o}rnle}, {L{\'o}pez-Caniego}, {Lubin}, {Mac{\'\i}as-P{\'e}rez}, {Maggio}, {Maino}, {Mandolesi}, {Mangilli}, {Marchini}, {Maris}, {Martin}, {Martinelli}, {Mart{\'\i}nez-Gonz{\'a}lez}, {Masi}, {Matarrese}, {McGehee}, {Meinhold}, {Melchiorri}, {Melin}, {Mendes}, {Mennella}, {Migliaccio}, {Millea}, {Mitra}, {Miville-Desch{\^e}nes}, {Moneti}, {Montier}, {Morgante}, {Mortlock},
  {Moss}, {Munshi}, {Murphy}, {Naselsky}, {Nati}, {Natoli}, {Netterfield}, {N{\o}rgaard-Nielsen}, {Noviello}, {Novikov}, {Novikov}, {Oxborrow}, {Paci}, {Pagano}, {Pajot}, {Paladini}, {Paoletti}, {Partridge}, {Pasian}, {Patanchon}, {Pearson}, {Perdereau}, {Perotto}, {Perrotta}, {Pettorino}, {Piacentini}, {Piat}, {Pierpaoli}, {Pietrobon}, {Plaszczynski}, {Pointecouteau}, {Polenta}, {Popa}, {Pratt}, {Pr{\'e}zeau}, {Prunet}, {Puget}, {Rachen}, {Reach}, {Rebolo}, {Reinecke}, {Remazeilles}, {Renault}, {Renzi}, {Ristorcelli}, {Rocha}, {Rosset}, {Rossetti}, {Roudier}, {Rouill{\'e} d'Orfeuil}, {Rowan-Robinson}, {Rubi{\~n}o-Mart{\'\i}n}, {Rusholme}, {Said}, {Salvatelli}, {Salvati}, {Sandri}, {Santos}, {Savelainen}, {Savini}, {Scott}, {Seiffert}, {Serra}, {Shellard}, {Spencer}, {Spinelli}, {Stolyarov}, {Stompor}, {Sudiwala}, {Sunyaev}, {Sutton}, {Suur-Uski}, {Sygnet}, {Tauber}, {Terenzi}, {Toffolatti}, {Tomasi}, {Tristram}, {Trombetti}, {Tucci}, {Tuovinen}, {T{\"u}rler}, {Umana}, {Valenziano}, {Valiviita}, {Van Tent},
  {Vielva}, {Villa}, {Wade}, {Wandelt}, {Wehus}, {White}, {White}, {Wilkinson}, {Yvon}, {Zacchei}, \& {Zonca}}]{Planck2016AA}
{Planck Collaboration}, {Ade}, P.~A.~R., {Aghanim}, N., {et~al.} 2016, \aap, 594, A13

\bibitem[{{Quinn} {et~al.}(1993){Quinn}, {Hernquist}, \& {Fullagar}}]{Quinn+93}
{Quinn}, P.~J., {Hernquist}, L., \& {Fullagar}, D.~P. 1993, \apj, 403, 74

\bibitem[{{Richstone}(1976)}]{Richstone1976ApJ}
{Richstone}, D.~O. 1976, \apj, 204, 642

\bibitem[{{Rodriguez-Gomez} {et~al.}(2016){Rodriguez-Gomez}, {Pillepich}, {Sales}, {Genel}, {Vogelsberger}, {Zhu}, {Wellons}, {Nelson}, {Torrey}, {Springel}, {Ma}, \& {Hernquist}}]{RodriguezGomez2016MNRAS}
{Rodriguez-Gomez}, V., {Pillepich}, A., {Sales}, L.~V., {et~al.} 2016, \mnras, 458, 2371

\bibitem[{{Rodriguez-Gomez} {et~al.}(2017){Rodriguez-Gomez}, {Sales}, {Genel}, {Pillepich}, {Zjupa}, {Nelson}, {Griffen}, {Torrey}, {Snyder}, {Vogelsberger}, {Springel}, {Ma}, \& {Hernquist}}]{Rodriguez2017MNRAS}
{Rodriguez-Gomez}, V., {Sales}, L.~V., {Genel}, S., {et~al.} 2017, \mnras, 467, 3083

\bibitem[{{Schnorr-M{\"u}ller} {et~al.}(2021){Schnorr-M{\"u}ller}, {Trevisan}, {Riffel}, {Chies-Santos}, {Furlanetto}, {Ricci}, {Lohmann}, {Flores-Freitas}, {Mallmann}, \& {Alamo-Mart{\'\i}nez}}]{Schnorr2021MNRAS}
{Schnorr-M{\"u}ller}, A., {Trevisan}, M., {Riffel}, R., {et~al.} 2021, \mnras, 507, 300

\bibitem[{{Shen} {et~al.}(2003){Shen}, {Mo}, {White}, {Blanton}, {Kauffmann}, {Voges}, {Brinkmann}, \& {Csabai}}]{Shen2003MNRAS}
{Shen}, S., {Mo}, H.~J., {White}, S. D.~M., {et~al.} 2003, \mnras, 343, 978

\bibitem[{{Silk} \& {Rees}(1998)}]{Silk1998}
{Silk}, J. \& {Rees}, M.~J. 1998, \aap, 331, L1

\bibitem[{{Spitzer} \& {Schwarzschild}(1951)}]{Spitzer&Schwarzschild51}
{Spitzer}, Lyman, {\rm Jr}. \& {Schwarzschild}, M. 1951, \apj, 114, 385

\bibitem[{{Springel}(2010)}]{Springel10}
{Springel}, V. 2010, \mnras, 401, 791

\bibitem[{{Springel} {et~al.}(2018){Springel}, {Pakmor}, {Pillepich}, {Weinberger}, {Nelson}, {Hernquist}, {Vogelsberger}, {Genel}, {Torrey}, {Marinacci}, \& {Naiman}}]{Springel2018MNRAS}
{Springel}, V., {Pakmor}, R., {Pillepich}, A., {et~al.} 2018, \mnras, 475, 676

\bibitem[{{Springel} {et~al.}(2001){Springel}, {White}, {Tormen}, \& {Kauffmann}}]{Springel2001MNRAS}
{Springel}, V., {White}, S. D.~M., {Tormen}, G., \& {Kauffmann}, G. 2001, \mnras, 328, 726

\bibitem[{{T\'oth} \& {Ostriker}(1992)}]{Toth&Ostriker92}
{T\'oth}, G. \& {Ostriker}, J.~P. 1992, \apj, 389, 5

\bibitem[{{van Dokkum} {et~al.}(2008){van Dokkum}, {Franx}, {Kriek}, {Holden}, {Illingworth}, {Magee}, {Bouwens}, {Marchesini}, {Quadri}, {Rudnick}, {Taylor}, \& {Toft}}]{vanDokkum+08}
{van Dokkum}, P.~G., {Franx}, M., {Kriek}, M., {et~al.} 2008, \apjl, 677, L5

\bibitem[{{Wang} {et~al.}(2023){Wang}, {Peng}, {Liu}, {Mihos}, {C{\^o}t{\'e}}, {Ferrarese}, {Taylor}, {Blakeslee}, {Cuillandre}, {Duc}, {Guhathakurta}, {Gwyn}, {Ko}, {Lan{\c{c}}on}, {Lim}, {MacArthur}, {Puzia}, {Roediger}, {Sales}, {S{\'a}nchez-Janssen}, {Spengler}, {Toloba}, {Zhang}, \& {Zhu}}]{Wang2023Nature}
{Wang}, K., {Peng}, E.~W., {Liu}, C., {et~al.} 2023, \nat, 623, 296

\bibitem[{{Watts} \& {Bekki}(2016)}]{Watts2016MNRAS}
{Watts}, A. \& {Bekki}, K. 2016, \mnras, 462, 3314

\bibitem[{{Weinberger} {et~al.}(2017){Weinberger}, {Springel}, {Hernquist}, {Pillepich}, {Marinacci}, {Pakmor}, {Nelson}, {Genel}, {Vogelsberger}, {Naiman}, \& {Torrey}}]{Weinberger2017MNRAS}
{Weinberger}, R., {Springel}, V., {Hernquist}, L., {et~al.} 2017, \mnras, 465, 3291

\bibitem[{{Wetzel} {et~al.}(2023){Wetzel}, {Hayward}, {Sand erson}, {Ma}, {Angl{\'e}s-Alc{\'a}zar}, {Feldmann}, {Chan}, {El-Badry}, {Wheeler}, {Garrison-Kimmel}, {Nikakhtar}, {Panithanpaisal}, {Arora}, {Gurvich}, {Samuel}, {Sameie}, {Pand ya}, {Hafen}, {Hummels}, {Loebman}, {Boylan-Kolchin}, {Bullock}, {Faucher-Gigu{\`e}re}, {Kere{\v{s}}}, {Quataert}, \& {Hopkins}}]{Wetzel+23}
{Wetzel}, A., {Hayward}, C.~C., {Sand erson}, R.~E., {et~al.} 2023, \apjs, 265, 44

\bibitem[{{Zolotov} {et~al.}(2015){Zolotov}, {Dekel}, {Mandelker}, {Tweed}, {Inoue}, {DeGraf}, {Ceverino}, {Primack}, {Barro}, \& {Faber}}]{Zolotov2015MNRAS}
{Zolotov}, A., {Dekel}, A., {Mandelker}, N., {et~al.} 2015, \mnras, 450, 2327

\end{thebibliography}

\end{document}